\documentclass{mn2e}        
\usepackage{graphicx}

\title[The iron lines from black-hole accretion discs]{On variability
and spectral distortion of the fluorescent iron lines from black-hole
accretion discs }

\author[A. Nied\'{z}wiecki and  P. T. \.{Z}ycki] {Andrzej
Nied\'{z}wiecki$^1$\thanks{e-mail: niedzwiecki@uni.lodz.pl}  and
Piotr T. \.{Z}ycki$^2$\\              
$^1${\L}\'{o}d\'{z} University, Department of Physics, Pomorska 149/153,   
90-236 {\L}\'{o}d\'{z}, Poland \\    
$^2$Nicolaus Copernicus Astronomical Center, Bartycka 18, 00-716 Warsaw, 
Poland}

\date{Accepted 2007 November 16. Received 2007 October 23; in original form 2007
January 12}
\pagerange{\pageref{firstpage}--\pageref{lastpage}}  \pubyear{}

\begin{document}

\maketitle

\label{firstpage}

\begin{abstract}
We investigate properties of iron fluorescent line arising as a result
of illumination of a black hole accretion disc by an X-ray source
located above the disc surface. We study in details the light-bending
model of variability of the line, extending previous work on the
subject.

We indicate bending of photon trajectories to the equatorial plane,
which is a distinct property of the Kerr metric, as the most feasible
effect underlying  reduced variability of the line observed in several
objects.  A model involving an X-ray source with a varying radial
distance, located within a few central gravitational radii around a
rapidly rotating black hole, close to the disc surface, may explain
both the elongated red wing of the line profile and the complex
variability pattern observed in MCG--6-30-15 by {\it XMM-Newton}.

We point out also that  illumination by radiation which returns to the
disc (following the previous reflection) contributes significantly to
formation of the line profile  in some cases.  As a result of this
effect, the line profile always has a pronounced blue  peak (which is
not observed in the deep minimum state in MCG--6-30-15), unless the
reflecting material is absent within the innermost 2--3 gravitational
radii.

\end{abstract}


\section{Introduction}

Broad iron lines observed from many black hole systems most likely
originate from the innermost regions of an accretion disc and their
profiles are shaped by gravitational redshift and Doppler shifts.
Modelling of the lines observed in several objects requires strongly
enhanced fluorescent emission from a few gravitational radii (e.g.,
Wilms  et al.~2001, Fabian et al.~2002, Miller et al.~2002, Miller et
al.~2004;  see review in Reynolds \& Nowak 2003), which in turn
indicates that  a primary source of hard X-ray emission must also be
located close to the  black hole.  Thus, both the primary and
reflected emission should be  subject to strong gravity effects.  
These effects are also tentatively considered as an explanation of the
complex variability  pattern characterising radiation reflected from
disc, including the iron  line, and the primary hard X-ray continuum
emission.  Namely, weak  variability of the reflected component,
uncorrelated with the variability  of the primary emission, has been
reported in a number of sources (e.g.,  Vaughan \& Fabian 2004; see
review in Fabian \& Miniutti 2005). This is contrary to expectations
from a  simple geometrical model of a hard X-ray source located close
to the  reflecting disc, where a strict correlation between variations
of the  primary and reflected emission should be observed.

Fabian \& Vaughan (2003)  first argued that such a reduced variability
may be explained by relativistic effects, in particular by light
bending and focusing the primary emission towards the accretion
disc. Qualitatively, variations of the reflected emission should be
much weaker as  changes of the height of the X-ray source cause
variations of its observed luminosity at infinity, while changes of
the flux received by the disc (enhanced by the gravitational focusing;
e.g., Matt et al.\ 1992; Martocchia \& Matt 1996; Petrucci \& Henri
1997) are much weaker.  However, for a static primary source located
on the symmetry axis, the illuminating radiation is focused into the
innermost part of the disc and the reflected emission is subject to
similar light bending as the primary.  As a result, similar
variability characterises the primary and reflected emission, at least
for observers with low inclination angles.
  
Miniutti et al.\ (2003) and Miniutti \& Fabian (2004) have developed
further  this model  to include rotation of the primary source around
the axis and the resulting beaming of its emission toward outer
regions of the disc.  Then, variations of the reflected emission are
reduced because of the more extended region it is produced.
Predictions of their model are found to be consistent with
observations of black-hole systems, e.g.\ by Miniutti et al.\ (2003),
Miniutti, Fabian \& Miller (2004), Fabian et al.\ (2004).  However, a
number of important effects were not systematically studied.  In
particular, a specific pattern of motion of a primary source is
assumed (including both location relative to the symmetry axis and
azimuthal motion) but it is not discussed how strongly the resulting
properties depend on these assumptions.  For example,  corotation of
the primary source with the disc is assumed by Miniutti  \& Fabian
(2004), which is likely  close  to the disc surface, but at most
approximate at high latitudes.  As the azimuthal motion of the primary
source relative to the disc may affect significantly the reflected
emission (see e.g.\ Reynolds \& Fabian 1997), it is not clear whether
predictions of this model are specific to the underlying assumptions
or generic to models involving the light bending.

In this paper we systematically analyse the light bending model,
concentrating on strong gravity effects. We neglect some other effects
which may contribute to the original variability problem, for example,
ionization of the disc surface (Nayakshin \& Kazanas 2002).  We focus
here on the iron K$\alpha$ line; an analysis of reflected emission
including the Compton reflected radiation will be presented in our
next paper.    We study in details a number  of geometrical scenarios
of the source location  and  motion. We point out certain inadequacies
in the  original computations of Miniutti \& Fabian (2004), and how
they influence  their quantitative results.  We find also a novel
scenario in which reduction of the line variability follows directly
from properties of photon transfer in the Kerr metric.  Namely, we
find that a source located close to the disc surface,  with a varying
radial distance from a Kerr black hole, gives rise to both an
approximately constant illumination of the  surrounding disc and  a
very strong variability of the  primary emission observed at
infinity. Some aspects of variability in similar models  have been
considered recently e.g.\ by Czerny et al.\ (2004) and Pech\'acek et
al.\  (2005), however, these studies considered only emission emerging
locally from the region under the source. On the other hand, we find
that transfer  of primary emission to more distant regions of the disc
is  crucial for variability effects  in such a model.
  
We concentrate on low inclination objects,  with application to
Seyfert 1 galaxies in mind. Obviously, the Doppler  shifts are more
pronounced at high inclinations, but observational studies  of high
inclination objects are less advanced, either because they are
obscured (as Seyfert 2 galaxies), or because the studies on dynamical
time  scale are not possible (as in stellar mass black hole systems).
Furthermore, we consider only  emission averaged  over at least an
orbital period. Effects resulting from varying azimuthal location of
an off-axis source, with respect to observer, are studied, e.g., in
Ruszkowski (2000), Yu \& Lu (2000) and Goyder \& Lasenby (2004).

In Section 3 we analyse various physical effects relevant to formation
of relativistic line profiles and variability effects; in Section 4 we
apply our results to a Seyfert 1 galaxy MCG--6-30-15 which is the best
studied object with clear signatures of strong gravity effects.

\section{Model description}

We consider an accretion disc, surrounding a Kerr black hole,
irradiated by  X-rays emitted from an isotropic point source
(hereafter referred to  as the source of primary emission).  The black
hole is characterised by its mass, $M$, and angular momentum, $J$. We
use the Boyer-Lindquist (BL) coordinate system $x^i =
(t,R,\theta,\phi)$.  We also make use of locally non-rotating (LNR)
frames (Bardeen, Press \& Teukolsky 1972), which are particularly
convenient for studying processes in the Kerr geometry (in the
Schwarzschild metric the LNR frame is equivalent to the rest frame of
a local static observer).

The following dimensionless parameters are used in the paper
\begin{equation}
r = {R \over R_{\rm g}},  ~~~a = {J \over c R_{\rm g} M}, ~~~\Omega =
{{\rm d} \phi \over {\rm d} \hat t},
\end{equation}
where  $R_{\rm g} = GM/c^2$ is the gravitational radius and $\hat t =
ct/R_{\rm g}$. In most cases we compute spectra for  $a=0.998$.  To
illustrate effects which are uniquely related to black hole rotation,
we compare in some cases spectra for $a=0.998$ with spectra for
corresponding models in the Schwarzschild metric ($a=0$).

We assume that a geometrically thin, neutral,  optically thick disc is
located in the equatorial plane of the Kerr geometry. The radial
coordinate  in the disc plane is denoted by $r_{\rm d}$. For distances
larger than the radius of the marginally stable circular orbit,
$r_{\rm ms}$  ($=1.23$ and 6 for $a=0.998$ and $a=0$, respectively) we
assume a circular motion of matter  forming the disc,  with Keplerian
angular velocity (Bardeen et al.\ 1972),
\begin{equation}
\Omega_{\rm K}(r_{\rm d}) = {1 \over a+r_{\rm d}^{3/2}}.
\end{equation}

We take into account emission from matter extending down to the event
horizon.  In most cases (except for model $S_0$, see below) we assume
that matter  free falls within $r_{\rm ms}$.  We strictly follow
Cunningham (1975) in description of velocity field  at $r_{\rm
d}<r_{\rm ms}$ in models assuming the free fall. In most cases  we
assume the outer radius of the emitting region in the disc $r_{\rm
out}=600$, but we compute also spectra for smaller $r_{\rm out}$ for
comparison with previous studies.

We assume that the primary X-ray emission is generated isotropically
in the source rest frame.  We neglect transversal or radial motion of
the source, $u^{\theta} = 0$ and $u^r = 0$, but take into account
azimuthal motion.   The motion is characterised by $V \equiv v^\phi /
c$, where $v^\phi$ is the  azimuthal velocity with respect to the LNR
frame.  To illustrate influence of $V$ we compare models with $V=0$,
which allows to separate effects due to gravity from kinematic
effects, and $V=0.5$, which is a typical value of orbital velocity in
the innermost regions.

Inclination of the rotation axis of the black hole to the line of
sight is given by $\mu_{\rm obs} \equiv \cos \theta_{\rm obs}$.  Most
results (except for these shown in Fig.~\ref{fig:f13})  presented in
this paper are for $\mu_{\rm obs}=0.85$, which is typical  for objects
presumably affected by strong gravity  (in particular, MCG--6-30-15
and Narrow Line Seyfert 1 galaxies).

\subsection{Monte Carlo}

We use a Monte Carlo method, involving a fully general relativistic
(GR)  treatment of photon transfer in the Kerr space-time, to find
spectra of primary emission and Fe line observed by a distant
observer.  A large number of photons are generated from the primary
source with isotropic distribution of initial directions in the source
rest frame.  Lorentz transformation, with relative velocity $V$,
yields photon energy, $E_{\rm ln}$, and momentum components (in the BL
coordinate directions), $p_{(r)}$, $p_{(\theta)}$  and $p_{(\phi)}$,
in the LNR frame.  These, in turn, yield constants of motion
(cf. eq.~(10) in Nied\'zwiecki (2005))
\begin{equation}
E_{\rm inf}  =  \left( {\Delta \Sigma \over A} \right)^{1/2} E_{\rm
ln} +  \left( {A \over \Sigma} \right)^{1/2} \omega \sin \theta
p_{(\phi)} c
\label{einf}
\end{equation}
\begin{eqnarray}
\lambda  \equiv  {L c \over E_{\rm inf} R_{\rm g}} & = & \left( {A
\over \Sigma} \right)^{1/2}  {\sin \theta p_{(\phi)}c \over E_{\rm
inf} } \label{lambda} \\     
\eta  \equiv  {Q c^2 \over E_{\rm inf}^2
R_{\rm g}^2} & = &   {p_{(\phi)}^2 c^2 \over  E_{\rm inf}^2} \Sigma +
\cos^2\theta  \left( {\lambda^2 \over \sin^2 \theta} - a^2 \right),
\label{constants}
\end{eqnarray}
where $E_{\rm inf}$ is the photon energy at infinity, $Q$ is the
Carter's constant, $L$ is the component of angular momentum parallel
to the black hole rotation axis, and
\[
\Delta=r^2 - 2r +a^2,~~~\Sigma= r^2 + a^2 \cos^2 \theta,
\]
\begin{equation}
A=(r^2 + a^2)^2 - a^2 \Delta \sin^2 \theta,~~~\omega = 2 a r /A.
\label{delsig}
\end{equation}
A photon trajectory is fully determined by $\lambda$ and $\eta$.

For each photon, equations of motion are solved to find whether the
photon crosses the event horizon, hits the disc surface or escapes
directly to distant observer.  Photon emission rate, yielding
normalisation of directly observed  and illuminating fluxes (see
below), is corrected by time dilation factor,
\begin{equation}
g_t \equiv {{\rm d} t_{\rm s} \over {{\rm d} t}} =  \left( {\Sigma
\Delta \over A} \right)^{1/2} \left( 1 - V^2 \right)^{1/2},
\label{dtdt}
\end{equation}
where ${\rm d} t_{\rm s}$ is the time interval in the source rest
frame.

For a photon hitting  the disc surface, at $r_{\rm d}$, we find
incidence angle and energy, $E_{\rm disc}$, in the disc rest frame.
Then, for $E_{\rm disc} > 7.1$ keV, we generate  an iron K$\alpha$
photon, with energy 6.4 keV, emerging from the disc.   The relative
weight of the Fe photon is related  to the initial energy and
direction of an incident photon by the  quasi-analytic formula
[eqs.~(4-6)] from George \& Fabian (1991).  We modify the original
formula by multiplying it by a factor of 1.3 to  account for different
elemental abundances (we use abundances of Anders \&  Grevesse 1989,
while George \& Fabian assumed abundances of Anders \&  Ebihara
1982). We have verified the factor of 1.3 using the Monte Carlo  code
of \.{Z}ycki \& Czerny (1994).  Initial direction of Fe K$\alpha$
photons is generated with uniform distribution of azimuthal angle but
we consider various distributions of the polar emission angle,
$\mu_{\rm em} \equiv \cos \theta_{\rm em}$.   In most cases we assume
a limb darkening in electron scattering limit, $I(\mu_{\rm em})
\propto 1 + 2.06 \mu_{\rm em}$ but in Sections \ref{sec:limb} and
\ref{sec:mcg} we take into account  other angular laws.

For photons hitting the disc we perform also simulation of Compton
reflection.  Then, we solve equations of motion in the Kerr metric for
both the Fe K$\alpha$ and Compton reflected photons emerging from
$r_{\rm d}$. The detailed analysis of the spectral  component formed
by the latter, reaching a distant observer, will be presented in our
next paper. Here we take into account effects due to reflected photons
which return to the disc giving rise to fluorescence (Section
\ref{sec:bend}).  Furthermore, in Section \ref{sec:ew} we take into
account the Compton reflected component for computing the equivalent
width of the Fe line.

\subsection{The models}

We study properties of the iron line, in particular effects related to
azimuthal motion and returning radiation, for two extreme locations of
the primary source,  represented by models $A$ and $S$, defined below,
with the source located close to the symmetry axis and close to disc
surface, respectively.  In these models, location of the primary
source is given by its BL coordinates, $r_{\rm s}$ and  $\theta_{\rm
s}$. We illustrate effects resulting from varying distance of the
source by changing $r_{\rm s}$ at  constant $\theta_{\rm s}$, which
seems appropriate for the innermost part of the flow  (e.g.\ numerical
simulations typically show conical coronae or outflows rather than
plane parallel structures).  We consider also a recently popular
variability model with position of the primary source parametrised by
$h_{\rm s} \equiv r_{\rm s} \cos \theta_{\rm s}$, defined as model $C$
below, in which the polar position of the source covers the whole
range of values of $\theta_{\rm s}$.  Specifically, model $C$ is
equivalent   to model $S$ for small $h_{\rm s}$  and to model $A$ for
high $h_{\rm s}$.

\subsubsection{Model $A$}
In model $A$ we consider a primary source located close to the axis.
All model $A$ spectra in this paper correspond to $\theta_{\rm
s}=0.05$ rad.  Our model $A$ with $V=0$ is virtually equivalent to
models with on-axis source, e.g.\ Matt et al.~(1992), Martocchia \&
Matt (1996), Petrucci \& Henri (1997), Martocchia, Karas \& Matt
(2000).  Our assumption of location being slightly displaced from the
axis allows us to study the  impact  of azimuthal motion in the polar
region.

\subsubsection{Model $S$, $S_{\rm K}$, $S_0$}
In model $S$ we consider a source located close to the disc surface.
All model $S$ spectra shown in this paper correspond to $\theta_{\rm
s}=1.5$ rad, i.e.\ $h_{\rm s}=0.07 r_{\rm s}$.  We discuss the
dependence of model $S$ on $V$, but in Section \ref{sec:mcg}   we
focus on a scenario with the primary source rigidly coupled to the
underlying disc and corotating with Keplerian velocity, $\Omega_{\rm
K}(r_{\rm s})$. This specific case  is denoted as model $S_{\rm K}$.
For $a=0.998$ we consider only $r_{\rm s} > r_{\rm ms}$.  All
variability effects in this model are studied for varying distance,
$r_{\rm s}$,  and a constant polar angle $\theta_{\rm s}$.

 We also introduce another model, $S_0$, for a non-rotating black
hole, $a=0$, but assuming a quasi-Keplerian velocity of accreting
matter in the plunging region below $r_{\rm ms}$: $v^r=0$ and
$v^{\phi}=0.5c$   (instead of the usual free fall assumption). Such a
velocity field is very similar to the velocity field of an accretion
disc around an $a=0.998$  black hole, in the region $1.25 < r < 6$
(where $v_{\rm K}(r=3.5)=0.5c$; $v_{\rm K}$ increases from $0.4c$ at
$r=6$ to $0.57c$ at $ r \le 2$).  This model can thus serve as a
comparison model to study effects specific to space-time around
rapidly rotating black hole, which are significant for $r_{\rm s} <
6$.  We emphasise that we do not consider  this as a likely scenario,
we are not aware of any results supporting such a rotating flow within
$r_{\rm ms}$ [although e.g.\  Agol \& Krolik (2002) argue that distant
regions of the flow may be linked through magnetic tension, which
presumably could affect motion of the flow in the plunging region].
We consider model $S_0$ only to separate effects due to gravitational
energy shift and bending of trajectories from kinematic effects.

\subsubsection{Model $C$}

Following Miniutti \& Fabian (2004) we consider a model with a
cylindrical-like motion of the primary source, referred to as model
$C$, with the  source located at constant projected radial distance
$\rho_{\rm s} \equiv r_{\rm s}\sin \theta_{\rm s}$ and changing the
height, $h_{\rm s} $, above the disc surface.  Furthermore, model $C$
assumes that at each $h_{\rm s}$ the source has the same angular
velocity, equal to the Keplerian velocity at $\rho_{\rm s}$,
$\Omega_{\rm K}(\rho_{\rm s})$.  This yields the following velocity of
the source rest frame relative to the LNR frame (Bardeen et al.\ 1972)
\begin{equation}      
V  = { \sin{\theta} A \over \Sigma \Delta^{1/2} } \left(  \Omega_{\rm
   K} - \omega \right).
\label{v}
\end{equation}
All spectra for model $C$  shown in this paper are for $\rho_{\rm
s}=2$.

Note that equation (\ref{v}) yields, for constant angular velocity, a
peculiar dependence of $V$ on source location. Namely,  $V$ increases
with increasing $h_{\rm s}$ at small heights, achieves a maximum value
at $h_{\rm s} \approx \rho_{\rm s}$ and then decreases.  E.g.\ for
$\rho_{\rm s}=2$: $V(h_{\rm s}=0)=0.57$,  $V(h_{\rm s}=2)=0.69$  and
$V(h_{\rm s}=20)=0.55$. These changes of $V$, with $h_{\rm s} $,
appear  to be crucial for properties of model $C$. To illustrate
impact of kinematic  assumptions, we show in Section \ref{sec:var}
variability in a model with primary source changing $h_{\rm s}$ at
constant $\rho_{\rm s}$, as in model $C$, but with constant  angular
{\em momentum\/} equal to angular momentum on a Keplerian orbit in the
equatorial plane at $\rho_{\rm s}$, $l_{\rm K}$ [yielding $\Omega =
(g^{\phi t} - l_{\rm K} g^{\phi \phi})/ (g^{tt} - l_{\rm K}g^{t
\phi})$].

Suebsewong et al.\ (2006) study effects of changing $\rho_{\rm s}$  at
constant $h_{\rm s}$, keeping parametrisation of the source motion
introduced by Miniutti \& Fabian (2004). Again, we note that for small
$h_{\rm s}$, which are most relevant for such scenario, the assumed
pattern of motion yields velocity field strongly affecting the derived
properties.  E.g., for $h_{\rm s}=0.5$, $V = 0.6-0.8$ for $\rho_{\rm
s} \le 2$, i.e.\ achieves values much larger than those of a Keplerian
motion in the equatorial plane.  Furthermore, results of both Miniutti
\& Fabian (2004) and Suebsewong et al.\ (2006) for large $\rho_{\rm
s}$ or $h_{\rm s}$ are strongly affected by the assumed value of
$r_{\rm out}=100$, see Section \ref{sec:var:c}.

\subsection{Intrinsic luminosity}

In all models we assume that the primary emission has a power-law
spectrum, with a photon spectral  index $\Gamma =2$, and an
exponential  cut-off at $E_{\rm cut}=150$ keV [this cut-off energy is
(marginally) relevant for the fluorescent line only due to effect of
returning radiation].   Units for spectra, presented in Section 3, are
arbitrary but all have the same normalisation, with intrinsic
luminosity, $L_0$, yielding the energy  flux of primary emission, per
unit solid angle in the rest frame of the source, $F_{\rm E}$,
normalised to unity at 1 keV. Assumption of constant intrinsic
luminosity is released in two cases: (i) in model $S_{\rm K}^{\rm PT}$
which adopts kinematic assumption of model $S_{\rm K}$, but assumes
intrinsic luminosity changing with $r_{\rm s}$ according to
dissipation rate per unit area in a Keplerian disc (Page \& Thorne
1974); (ii) in models $\hat C$, $\hat S_{\rm K}$ and $\hat S_0$,
considered in Section \ref{sec:var},  which adopt assumptions of
models $C$, $S_{\rm K}$ and $S_0$, respectively, but neglect time
dilation at primary source, i.e.\ assume the source intrinsic
luminosity  $= L_0 /g_t$.

We present spectra averaged over azimuthal angle. Although in models
defined above the axial symmetry is broken,  the orbital periods  are
much shorter (with  one exception, see below) than  the timescale, 10
ks,  over which  variability properties are investigated in the
highest quality data.  Note that, for high $a$, azimuthal averaging is
justified also  for a source with $V=0$, which rotates with respect to
a distant observer with  angular velocity $\omega$, comparable to
$\Omega_{\rm K}$ at small $r$, e.g.\ $\omega \approx 0.2\Omega_{\rm
K}$ at $r = 4.5$.

For a black hole with $M=10^7 M_{\sun}$ the orbital period exceeds 10
ks in model $S_{\rm K}$ with $r_{\rm s} > 10$. Indeed, narrow,
slightly shifted  lines observed is some objects with such large
masses are considered to arise from localised  spots on the disc (e.g.
Dov\v{c}iak et al.~2004b) in a model equivalent to our $S_{\rm K}$
with large $r_{\rm s}$. Note, however, that this range of parameters
is not relevant to effects considered in our paper. In particular, GR
effects are weak and effects of black hole rotation are completely
negligible for such distances. We  analyse variability effects
averaged over an orbital period for such large $r_{\rm s}$ for
potential applicability to lower mass systems.

Finally, we ignore time delays between the primary and reprocessed
radiation, related  to light-travel times between the source and
various parts of the disc. For a central  source illuminating the ring
of radius $R_{\rm d}(\gg R_{\rm g})$, these delay times  are between
$(1 - \sin \theta_{\rm obs})R_{\rm d}/c$ and $(1 + \sin \theta_{\rm
obs})R_{\rm d}/c$.  For $\theta_{\rm obs}=30 \degr$, the minimum and
maximum delay is $2.5 \times 10^{-6}  r_{\rm d} M/M_{\sun}$ s and $7.5
\times 10^{-6} r_{\rm d} M/M_{\sun}$ s, respectively. Then, for
$M=10^7 M_{\sun}$ and $r_{\rm d} > 150$, the delays exceed  typical
times, $\ge 10$ ks,  for high quality spectra.

We emphasise  that in most cases these delays are not important to
variability effects studied here, regardless  of the value of $M$, as
the major part of reprocessed radiation originates from $r_{\rm d} \la
10$ and the fraction of Fe photons coming from $r_{\rm d} > 100$ is
smaller than 1 per cent.  Significant irradiation of disc at $r_{\rm
d} > 100$ occurs only for $h_{\rm s} > 10$ in model $C$.  In this case
the delays may affect variability for large $M$, we discuss this issue
further in Section \ref{sec:var:c}.  We choose  $r_{\rm out}=600$
because for this value properties of neither of our models  are
affected by neglecting reflection from distances larger than $r_{\rm
out}$. Even in the most extreme case of model $C$ with $h_{\rm s} =
20$ contribution from $r_{\rm d}>600$ would not exceed 1 per cent.
However, in other models a smaller value, e.g. $r_{\rm out} = 100$,
could be safely assumed.

\section{Results}

\subsection{Radial emissivity of Fe K$\alpha$ line.}

\begin{figure}
\centerline{\includegraphics[height=90mm]{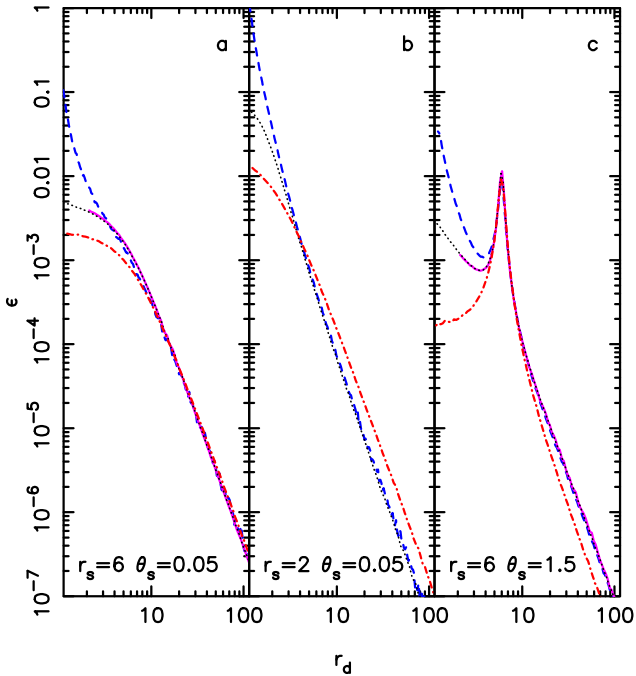}}
\caption{Radial illumination ($\epsilon_{\rm ph}$) and Fe K$\alpha$
emissivity ($\epsilon_{\rm Fe}$) profiles, averaged over azimuthal
angle, for a disc illuminated by a point source, with $V=0$: {\bf (a)}
model $A$ with  $r_{\rm s}=6$; {\bf (b)} model $A$ with   $r_{\rm
s}=2$; {\bf (c)} model $S$ with $r_{\rm s}=6$.    The dotted (black)
and dot-dashed (red) curves show $\epsilon_{\rm ph}$  in the Kerr
(with $a=0.998$) and flat space-time, respectively. The solid
(magenta) curves in panels (a) and (c), approximately coincident with
the dotted curves at $r_{\rm d}>2$, show $\epsilon_{\rm ph}$ for the
Schwarzschild space-time.  The dashed (blue) curves show
$\epsilon_{\rm Fe}$ for the Kerr metric (and photon spectral index
$\Gamma = 2$). The illuminating photon flux is normalised by total
number of photons emitted from the primary source.  The $\epsilon_{\rm
Fe}$ profiles are normalised to dotted curves at large $r_{\rm d}$.
Deviations between dot-dashed and dotted curves illustrate impact of
GR effects  on the radial illumination pattern. Deviations between
dotted and dashed curves are primarily due to blueshift of photons in
the disc rest frame.  }
\label{fig:f1}
\end{figure}

Fig.\ \ref{fig:f1} shows radial illumination profiles, $\epsilon_{\rm
ph}$,  and corresponding  radial emissivities  of Fe photons,
$\epsilon_{\rm Fe}$, resulting from illumination of the disc surface
by a point source.  The $\epsilon_{\rm ph}$ and $\epsilon_{\rm Fe}$
are given  in number of photons, per unit time and unit area of the
disc, hitting the disc surface and emerging   from the disc,
respectively, at distance $r_{\rm d}$. All radial profiles are
averaged over the azimuthal angle, $\phi$.

Note that the illuminating flux as well as photon emission rate are
equal when expressed both in the BL coordinate frame and in the disc
rest frame
\begin{equation}
\epsilon \equiv {N_{\rm ph} \over {\rm d} \hat t {\rm d} \hat S} =
{N_{\rm ph} \over {\rm d} t_{\rm d}  {\rm d} S_{\rm d}},
\end{equation}
where $N_{\rm ph}$ is the number of photons, ${\rm d} \hat t$ and
${\rm d} t_{\rm d}$ is the time interval in BL and disc rest frame,
respectively, and ${\rm d} \hat S$ and ${\rm d} S_{\rm d}$ is the
surface area element in the BL and disc rest frame, respectively.
Namely, the following relations ${\rm d} S_{\rm d} = \gamma
(A/\Delta)^{1/2} r^{-1}{\rm d} \hat S$ (e.g. Petrucci \& Henri 1997)
and ${\rm d} t_{\rm d} = \gamma^{-1} r (\Delta/A)^{1/2}  {\rm d} \hat
t$ (cf.~equation (\ref{dtdt}); $\gamma$ is the Lorentz factor
corresponding to Kepler velocity, $v_{\rm K}$, of disc element with
respect to the LNR frame) imply that both GR and special relativistic
(SR) effects cancel out.  Note also that (Lorentz) transformation
between the disc rest frame and the LNR frame (labelled by 'ln'
subscript)  includes additional time retardation terms,  ${\rm d}
S_{\rm ln} = \gamma^{-1}(1 - \mu v_{\rm K}/c)^{-1}{\rm d} S_{\rm d}$
and  ${\rm d} t_{\rm ln} = (1 - \mu v_{\rm K}/c) \gamma {\rm d} t_{\rm
d}$ (where $\mu$ is cosine of the angle, in disc rest frame, between
the direction of motion of a surface element and the direction of
emitted/incident photon), which again cancel out (Mathews 1982, Laor
\& Netzer 1990), yielding  ${\rm d} t_{\rm ln}{\rm d} S_{\rm ln} =
{\rm d} t_{\rm d}{\rm d} S_{\rm d}$ (the same conclusion follows
trivially from Lorentz invariance of the space-time volume  between
the LNR frame and the disc rest frame; Petrucci \& Henri 1997).

For clear illustration of effects resulting from light bending,  we
compare in Fig.\ \ref{fig:f1} the $\epsilon_{\rm ph}$ profiles for a
black hole accretion disc and a disc in flat space-time, with the same
position of the primary  source when expressed in BL and spherical
coordinates, respectively.  The profiles for a static point source in
flat space-time are determined by obvious geometrical effects
involving the distance from the primary source to disc element and
projection of the disc area, yielding
\begin{equation}
\epsilon_{\rm ph}(r_{\rm d}) \propto \int_0^{2\pi} [ r_{\rm d}^2 +
h_{\rm s}^2 + \rho_{\rm s}^2 - 2 r_{\rm d} \rho_{\rm s} \cos (\phi)
]^{-3/2} {\rm d} \phi,
\end{equation}
where $\rho_{\rm s}$ ($=r_{\rm s}\sin \theta_{\rm s}$)  is the
distance of the source from the symmetry axis, $h_{\rm s}$ ($=r_{\rm
s}\cos \theta_{\rm s}$) is the height of the source above the disc
surface  and $\phi$ is an azimuthal angle in the disc plane.  For
$\rho_{\rm s} \ll h_{\rm s}$, Figs.\ \ref{fig:f1}(a)(b), the profile
is flatter in the central region (at $r_{\rm d} < h_{\rm s}$) and
steepens to $\epsilon_{\rm ph} \propto r_{\rm d}^{-3}$  at $r_{\rm d}
\gg h_{\rm s}$. For $\rho_{\rm s} > h_{\rm s}$, Fig.\ \ref{fig:f1}(c),
the profile has a steep local maximum at $r_{\rm d} \approx \rho_{\rm
s}$.

In the black-hole model,  light bending enhances illumination of the
central region. For a source close to the symmetry axis the deflection
of light gives rise to $\epsilon_{\rm ph}$ significantly steeper, in
the innermost few $R_{\rm g}$, than in the flat model.  Note, however,
that only $r_{\rm s} \la 2$, Fig.\ \ref{fig:f1}(b),  leads to steep
illumination profile, with $q \ga 3$, where by $q$ we denote index of
a power-law approximating locally the radial profiles, i.e.\ $\epsilon
\propto r_{\rm d}^{-q}$.  For a source located close to the disc
surface, Fig.\ \ref{fig:f1}(c), deflection of  photon trajectories
increases illumination of the opposite (to the source) side of the
disc, enhancing $\epsilon_{\rm ph}$ both at $r_{\rm d}<r_{\rm s}$ and
(moderately) at $r_{\rm d}> r_{\rm s}$.

Dashed curves in Fig.\ \ref{fig:f1} show emissivity of Fe K$\alpha$
photons for the model with $a=0.998$. Within $r_{\rm d} \la 4$, the
$\epsilon_{\rm Fe}$ profile  is steeper than $\epsilon_{\rm ph}$. This
effect is primarily due to the gravitational blueshift of photons
hitting the disc surface, which increases the number of photons with
energies above the photoelectric threshold.  Note, however, that very
steep ($q \ga 4$) $\epsilon_{\rm Fe}$ occurs only in the part of the
disc extending within the ergosphere [the ergosphere is the region
contained within the surface given by $r_{\rm erg} = 1+(1 - a^2 \cos^2
\theta)^{1/2}$ ($\approx 2$ for $a=0.998$ and $\theta=\pi/2$)] where
the gravitational blueshift is sufficiently high, e.g.,   $E_{\rm
disc}/E_{\rm s}>4$ at $r_{\rm d}=2$ and $>10$ around  $r_{\rm
d}=1.2$($\approx r_{\rm ms}$), where $E_{\rm s}$ is the photon energy
in the source rest frame.

\subsection{Bending to the equatorial plane}
\label{sec:bend}

\begin{figure}
\centerline{\includegraphics[height=80mm]{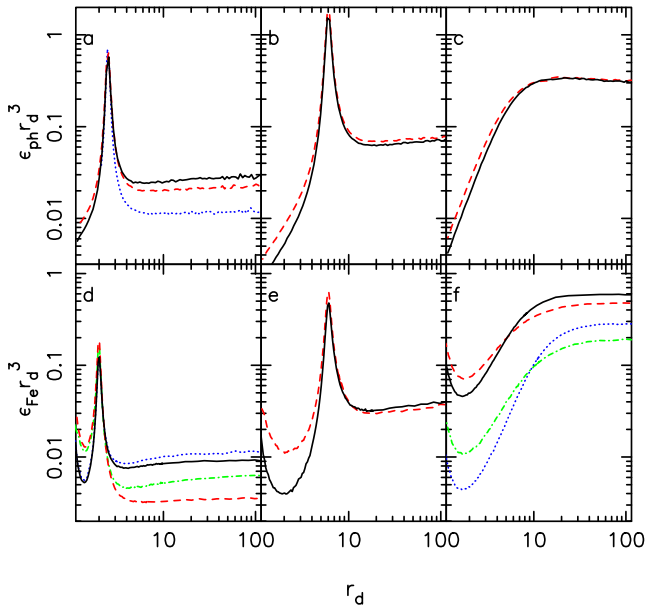}}
\caption{The figure illustrates dependence of $\epsilon_{\rm ph}$
(panels a-c) and $\epsilon_{\rm Fe}$ (panels d-f) on azimuthal
velocity of the primary source.  Panel (a) illustrates also bending to
equatorial plane in the Kerr metric by comparing $\epsilon_{\rm ph}$
for $a=0$ and $a=0.998$.   Panel (d) shows effect of returning
radiation on $\epsilon_{\rm Fe}$ for $V=0$ and $V=0.5$.  All
$\epsilon_{\rm ph}$ and $\epsilon_{\rm Fe}$, except for dotted (blue)
curve  in panel (a), are for $a=0.998$.  Panels (a-c) show
$\epsilon_{\rm ph}$ resulting from direct illumination (i.e.
neglecting returning radiation) for a source, with $V=0.5$ (solid
curves, black) and $V=0$ (dashed curves, red): {\bf (a)} model $S$
with $r_{\rm s}=2.5$; {\bf (b)} model $S$ with $r_{\rm s}=6$; {\bf
(c)} model $A$ with $r_{\rm s}=6$. The dotted (blue) curve in panel
(a) is for model $S$ with $a=0$, $V=0$ and $r_{\rm s}=2.5$.  {\bf (d)}
Model $S$ with $r_{\rm s}=2$. The solid (black) and dashed (red)
curves are for $V=0.5$ and $V=0$, respectively, and result from direct
illumination; the dotted (blue) and dot-dashed (green) curves are for
$V=0.5$ and $V=0$, respectively, and take into account the returning
radiation. {\bf (e)} Model $S$ with $r_{\rm s}=6$. The solid (black)
and  dashed (red) curves are for $V=0.5$ and 0, respectively. {\bf
(f)} The solid (black) and  dashed (red) curves are for model $A$,
$r_{\rm s}=6$, with $V=0.5$ and $V=0$, respectively.  The dotted
(blue) and dot-dashed (green) curves are for model $C$, $h_{\rm s}=8$,
with $V \approx 0.6$ [resulting from equation (\ref{v})] and $V=0$,
respectively.    }
\label{fig:f2}
\end{figure}

Trajectories of photons emitted from a source located at $r_{\rm s} >
4$ are  bent toward the centre, which effect operates in the same way
regardless of the black hole spin value.  As a result, identical
profiles of $\epsilon_{\rm ph}$ (at $r_{\rm d}>2$) are achieved for
high and low value of $a$, see Figs.\ \ref{fig:f1}(a)(c), in models
with $r_{\rm s}>4$.  For a source located within $r_{\rm s}<4$ from a
rapidly rotating black hole, an effect unique for the Kerr metric
operates, namely the emission is strongly focused  along the
equatorial plane (Cunningham 1975, Dabrowski et al.\ 1997).  This
property of the Kerr space-time leads to significant enhancement of
illumination of  the disc lying in its equatorial plane, particularly
for a source located within  the ergosphere.  Strength of the bending
decreases with increasing $r_{\rm s}$, but for $2<r_{\rm s} \la 4$
this effect remains crucial, e.g.\ a source with $V=0$ located at
$r_{\rm s}=2.5$ yields $\epsilon_{\rm ph}$ by a factor of 2 higher for
$a=0.998$ than for $a=0$, see Fig.\ \ref{fig:f2}(a). Crucially for the
strength of the Fe line observed by low-inclination observers, bending
to the equatorial plane enhances  irradiation of disc at $r_{\rm d} >
r_{\rm s}$, while bending toward the centre, which is a dominant
effect for higher $r_{\rm s}$,  enhances  irradiation at $r_{\rm d} <
r_{\rm s}$.

Deflection of photon trajectories toward the equatorial plane affects
both the primary emission and radiation reflected from inner parts of
the disc (see Cunningham 1976).  The former case is crucial for
properties of model $S$ as we discuss in Section \ref{sec:var}.   The
latter results in redistribution of illuminating flux on the disc
surface and in some cases yields significant increase of the Fe line
flux, as we show in this section.

\begin{figure}
\centerline{\includegraphics[height=42mm]{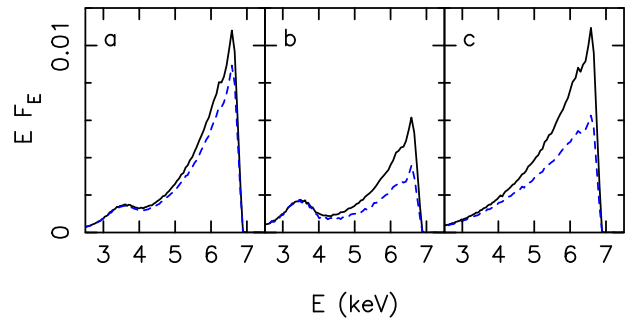}}
\caption{Effect of returning radiation on profiles of Fe K$\alpha$
line  for $\mu_{\rm obs} = 0.85$.  The dashed  (blue) profiles result
from direct illumination only, the solid (black) profiles take into
account radiation returning.    {\bf (a)}  Model $S$ with $V=0.5$ and
$r_{\rm s}=2$.   {\bf (b)}   Model $S$ with $V=0$ and $r_{\rm s}=2$.
{\bf (c)}   Model $A$ with $V=0$ and $r_{\rm s}=2$.    }
\label{fig:f3}
\end{figure}

Enhancement of Fe K$\alpha$ line by returning radiation was studied by
Dabrowski et al.\ (1997) in a model with a source corotating with the
disc, consistent with our model $S_{\rm K}$.  They find that the
returning continuum photons increase the flux of the line by no more
than 20 per cent.  In generic agreement with Dabrowski et al.\ (1997),
we find in model $S$ with $V=0.5$ (and similarly in model  $S_{\rm
K}$), with small $r_{\rm s}$, that returning continuum photons
increase $\epsilon_{\rm Fe}$ at  $r_{\rm d}> r_{\rm s}$  by $\ga 10$
per cent, see Fig.\ \ref{fig:f2}(d), yielding enhancement of the line
by the same order of magnitude,  see Fig.\ \ref{fig:f3}(a).  This
property is obvious, as (i) the reflecting albedo around the Fe
K$\alpha$  edge is about 10 per cent  and (ii) transfer of photons
from both the corotating source and the disc under the source, to
other regions of the disc, is similar. The overall effect of the
returning radiation has a minor impact on model involving primary
source corotating with the disc (see also Fig.~\ref{fig:f8}(b) below).

\begin{figure}
\centerline{\includegraphics[height=52mm]{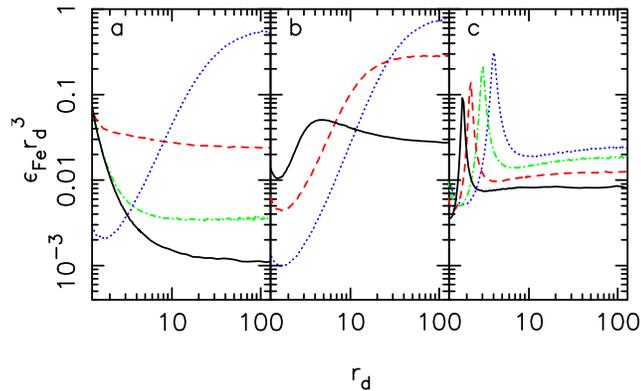}}
\caption{Radial $\epsilon_{\rm Fe}$ profiles resulting from direct
illumination in our basic models.   {\bf (a)} Model $A$ with $V=0$ and
$r_{\rm s}=1.6$ (solid curve, black),  3 (dashed curve, red)  and 20
(dotted curve, blue).  {\bf (b)} Model $C$ with $h_{\rm s}=2$ (solid
curve, black),  8 (dashed curve, red) and 20  (dotted curve, blue).
{\bf (c)}  Model $S_{\rm K}$ with $r_{\rm s} = 1.8$ (solid curve,
black), 2.2 (dashed curve, red), 3 (dot-dashed curve, green)  and 4
(dotted curve, blue).  The dot-dashed (green) curve in panel (a) takes
into account contribution of returning radiation  for $r_{\rm s}=1.6$.
}
\label{fig:f4}
\end{figure}

However, we note that a significantly stronger enhancement of the line
by returning photons may occur if the source does not corotate with
the disc.  Equations (\ref{einf}) and (\ref{lambda}) yield
\begin{equation}
E_{\rm inf} = { E_{\rm ln} \over 1 - \omega \lambda},
\end{equation}
which implies that the gravitational shift of photon energy between
the source, $r_{\rm s}$, and the disc, $r_{\rm d}$,
\begin{equation}
{ E_{\rm ln}(r_{\rm d}) \over E_{\rm ln}(r_{\rm s}) } = { 1 - \lambda
\omega(r_{\rm d}) \over 1 - \lambda \omega(r_{\rm s}) }.
\end{equation}
increases with decreasing $\lambda$ for $r_{\rm s} > r_{\rm d}$.
Then, the average gravitational  blueshift is higher for lower $V$
(implying on average lower $\lambda$ of emitted photons).
Furthermore, for model $S$ with $V=0$, Doppler blueshift of
illuminating photons is high due to velocity gradient between the
source and the disc. Combination of the above two effects results in
higher blueshift, for smaller $V$,  of  photons illuminating disc
within the ergosphere and thus  in the increase of magnitude of
radiation reflected from $r_{\rm d} < 2$ -   most of which  returns to
the disc, enhancing $\epsilon_{\rm Fe}$ more efficiently than for
$V=0.5$.  E.g.\ in model $S$ with  $r_{\rm s} = 2$ and $V=0$, the
resulting enhancement of $\epsilon_{\rm Fe}$ [Fig.~\ref{fig:f2}(d)]
and of the   line flux [Fig.\ \ref{fig:f3}(b)]  exceeds $50$ per cent.
Even more significant effect occurs for a source with a very centrally
concentrated (and weak beyond a few $R_{\rm g}$) direct illumination
of the disc, specifically, for a source with  $V=0$  at a small
$r_{\rm s}$ in model $A$, see Figs.\ \ref{fig:f4}(a) and
\ref{fig:f3}(c). E.g.\ for $r_{\rm s}=1.6$, due to contribution of
returning radiation, the line flux increases by a factor of $\approx
3$.

Enhancement of the Fe line by returning radiation is efficient only
for small $r_{\rm s}$.  For $r_{\rm s}>3$, the flux of returning
radiation is in general much smaller than the directly illuminating
flux.  Note also that all effects discussed in this section are unique
for high $a$ models. Specifically, we find that for $a=0$ returning
radiation enhances the Fe line flux at most by  a few per cent.

\subsection{Azimuthal motion}
\label{sec:vel}

We point out that azimuthal motion of the primary source affects
strongly the line flux through combination  of GR and SR effects.  The
relevant SR effects are discussed by Reynolds \& Fabian
(1997). Doppler shifts and aberration of the primary emission
influence both the number and average direction of primary photons
above the iron photoelectric threshold in the disc rest frame.  On the
other hand, azimuthal motion reduces slightly both the illuminating
and observed flux due to Lorentz time dilation (by $1/\gamma \approx
0.87$ for $V=0.5$).

Additional effects occur for a black hole accretion disc, most
pronouncedly  for small $r_{\rm s}$ ($<4$) and high $a$. With
increasing $V$, collimation toward direction of motion ($+\phi$
direction) increases fraction of photons bent  toward the equatorial
plane and thus enhances $\epsilon_{\rm ph}$. For $2 < r_{\rm s} \la
4$, this effect is moderate, e.g. in model $S$ with $r_{\rm s}=2.5$
increase of velocity from $V=0$ to $V=0.5$ results in increase of
$\epsilon_{\rm ph}$ at $r_{\rm d}>3$ by 20 per cent,  see Fig.\
\ref{fig:f2}(a). Much more efficient increase of the fraction of
primary photons bent to the disc - with increasing velocity - occurs
for  $r_{\rm s} \la 2$.  Furthermore, the SR collimation increases the
average $p_{(\phi)}$ and therefore the average energy of primary
photons increases, see equation (\ref{einf}), yielding additional
enhancement of the ionizing flux of photons hitting the disc.
Combination of these effects results in a strong enhancement of
$\epsilon_{\rm Fe}$  by increasing $V$.  E.g., in model $S$ with
$r_{\rm s}=2$, increase of $V$ from $V=0$ to $V=0.5$ results in the
increase of $\epsilon_{\rm Fe}$ by a factor of 2.5, Fig.\
\ref{fig:f2}(d).

\begin{figure}
\centerline{\includegraphics[height=80mm]{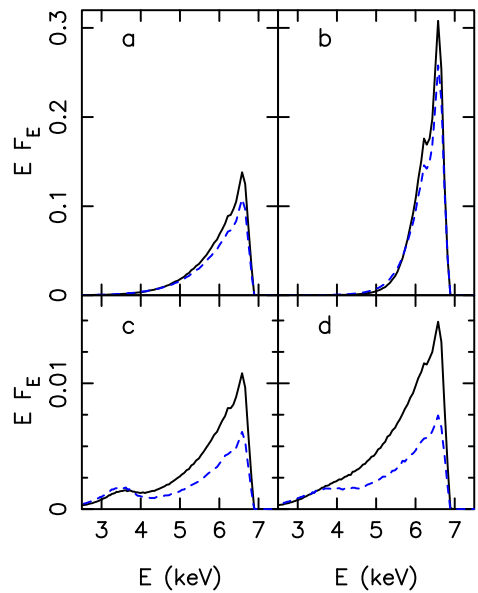}}
\caption{Impact of the primary source azimuthal velocity on the Fe
K$\alpha$ line profiles for $\mu_{\rm obs} = 0.85$. Panels (a) and (c)
compare profiles for $V=0.5$ (solid, black)  and $V=0$ (dashed, blue)
in model $A$ with $r_{\rm s}=6$ (a) and in model $S$ with $r_{\rm
s}=2$ (c). Panels (b) and (d) are for model $C$ with $h_{\rm s}=20$
(b) and $h_{\rm s}=0.5$ (d); the solid (black) profiles are for $V$
resulting from equation (\ref{v}), i.e.\ $V=0.55$ (b) and $V=0.59$
(d); the dashed (blue) profiles are for the same locations of primary
source but with $V=0$.  All profiles include contribution of returning
radiation.  }
\label{fig:f5}
\end{figure}

Note that the enhanced irradiation of the disc by a source with higher
$V$, discussed above,  results only from the direct emission from the
source.  On the other hand,  models with smaller $V$ are characterised
by stronger contribution of returning radiation, as discussed in
Section 3.2. Figs.\ \ref{fig:f5}(c)(d) show example profiles of lines
for $V=0$ and $V \simeq 0.5$, in models with $r_{\rm s} \approx 2$.

For $r_{\rm s}>4$ changes of the line caused by change of $V$ are more
subtle, see Figs.~\ref{fig:f5}(a)(b), and result mostly from the
Doppler blueshift.  As shown in Figs.~\ref{fig:f2}(b)(c), increase of
$V$ approximately does not change $\epsilon_{\rm ph}$ at $r_{\rm
d}>r_{\rm s}$  (assuming that the photon emissivity rate in the source
rest frame does not change). Although the SR collimation of primary
emission into the direction of motion  increases the fraction of
photons illuminating the disc at $r_{\rm d} > r_{\rm s}$, this
increase is balanced by the SR time dilation. On the other hand,
increase of $V$ reduces $\epsilon_{\rm ph}$ in the central region of
the disc, more noticeably for small $h_{\rm s}$ (compare
Fig.~\ref{fig:f2}(b) and Fig.~\ref{fig:f2}(c)), which effect is,
however, not important for face-on observers. Then, for high $r_{\rm
s}$, dependence of $\epsilon_{\rm Fe}$  (and hence of the emitted
line) on $V$ is mostly due to Doppler shifts. These energy shifts are
slightly higher for increasing $h_{\rm s}$ due to higher, on average,
velocity gradient between the source and the disc.

In general, the above GR and SR effects enhance the illuminating flux
at the expense of the primary flux observed at low inclinations, which
increases sensitivity of variability models (Section 3.5) on
assumptions regarding the azimuthal motion.  As in Section 3.2, we
emphasise that the GR effects discussed in this section  are specific
for models involving a rapidly rotating black hole.

\begin{figure}
\centerline{\includegraphics[height=64mm]{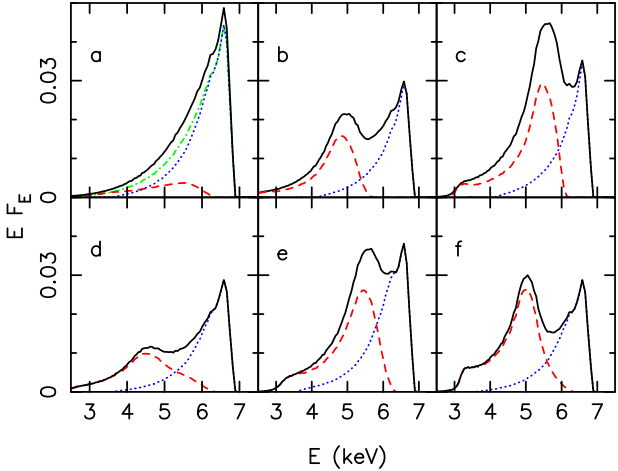}}
\caption{Impact of the space-time metric on the Fe K$\alpha$ line
profiles for $\mu_{\rm obs} = 0.85$. {\bf (a)} Model $A$, with $V=0$
and  $r_{\rm s}=4$. The solid (black) and dot-dashed (green) curve is
for $a=0.998$ and $a=0$, respectively.  Panels (b--f) are for models
$S$ with $V=0.5$: {\bf (b)} $r_{\rm s}=3$, $a=0.998$; the dashed (red)
curve shows contribution of photons emitted from a narrow ring
($r_{\rm d}=3.5-4.5$) under the source, the dotted (blue) curve shows
contribution of photons emitted from $r_{\rm d}>7$;  {\bf (c)} $r_{\rm
s}=4$, $a=0.998$; the dashed (red) and dotted (blue) curve shows
contribution of photons emitted from $r_{\rm d}=4.5-5.5$ and $r_{\rm
d}>7$, respectively; {\bf (d)}  $r_{\rm s}=4$, $a=0$ (free fall within
$r_{\rm ms}$); {\bf (e)}  $r_{\rm s}=5$, $a=0$ (free fall within
$r_{\rm ms}$); {\bf (f)} Model $S_0$ with $r_{\rm s}=4$.   Profiles
for $a=0$ in panels (a,d,e,f) are decomposed  into contribution from
the region of the disc within  $r_{\rm ms}$ (dashed, red) and beyond
$r_{\rm ms}$ (dotted, blue).  }
\label{fig:f6}
\end{figure}

\subsection{The red wing}
\label{sec:wing}

In this section we focus on the shape of the red wing (i.e.\ $<6$ keV)
of the line, which is formed mostly by photons emitted within a few
($\la 6$--7) $R_{\rm g}$. We consider effects of space-time metric on
the red wing shape and we indicate properties of model $S$ which
uniquely allow to investigate  effects related to value of $a$.
Applicability of various models  to modelling  the elongated wing
observed in MCG--6-30-15 is discussed in Section \ref{sec:mcg:profil}.

Profiles emerging due to combined emission from various $r_{\rm d}$
have qualitatively similar shapes for low and high $a$, see Fig.\
\ref{fig:f6}(a)].   The smooth low-energy tails of such line profiles
indicate that photons forming them come from a region at a few $R_{\rm
g}$ but they do not possess features allowing to  constrain more
detailed properties of the emitting region,  specifically the
space-time metric.   In particular, change of the angular law,
$I(\mu_{\rm em})$, has at least the same influence on the line shape
as a change of the value of $a$, compare Figs.\ \ref{fig:f6}(a) and
\ref{fig:f16}. Then, it has been argued  (e.g.\ Dov\v{c}iak, Karas \&
Yaqoob 2004a) that these time-averaged line profiles do not contain
sufficient information to yield unambiguous information about the
black hole spin.  Such a featureless wing is always produced as a
result  of illumination  by a source close to the axis, i.e.\ in our
model $A$ and similarly in model $C$ (except for very small $h_{\rm
s}$, see discussion of model $S$ below).

Detection of photons with energies $E_{\rm obs} < 4$ keV, i.e.\
redshifted by  $E_{\rm obs}/E_{\rm disc} < 0.6$, is the key property
which could be used to infer a high value of $a$ (for small
inclinations). Presence of such photons in the Fe K$\alpha$ line may
distinguish high from low values of $a$, under the assumption that in
the latter case the Fe photons emerge only from $\ge r_{\rm ms}$.  If
contribution from within  $r_{\rm ms}$ is taken into account, profiles
in models with $a=0$ and $a=0.998$ extend down to similar energies,
see Fig.\ \ref{fig:f6}(a). Models of the plunging region (e.g. Young,
Ross \& Fabian 1998) assess full ionization at $r_{\rm d}<5$ for
$a=0$, ruling out contribution from that region.  However,  they make
several simplifying assumptions, most importantly, they  base on
properties of a test particle motion.

Regardless of these details,  escape probability from the region
producing photons observed at $<4$ keV is small.  Therefore, even for
a source at a small height on the symmetry axis,  the resulting Fe
flux below 4 keV is rather weak, see the solid profile  in Fig.\
\ref{fig:f6}(a), and it may be difficult  to deconvolve this emission
from continuum components.  Generation  of a pronounced wing at such
energies requires a source located close to the disc surface in the
region emitting photons with the most extreme redshifts (see also
Section \ref{sec:mcg:profil}).  We point out that effects of
space-time metric can be studied  in line profiles emerging in such
cases, specifically, for a source located at (i) $r_{\rm s} \la 4$ and
(ii) close to the disc surface ($h_{\rm s} \le 0.1r_{\rm s}$). In such
a case, most of fluorescent photons are emitted from a relatively
narrow ring,  with the width $\Delta r_{\rm d} \la 1$, under the
primary source, see Fig.\ \ref{fig:f4}(c).  Emission from this ring
has a well defined gravitational redshift as well as Doppler shifts.
For $\mu_{\rm obs} = 0.85$, magnitude of  the gravitational redshift
($E_{\rm inf}/E_{\rm LN} < 0.7$ for $r_{\rm d}<4$) exceeds the Doppler
blueshift and photons emitted from $r_{\rm d} \approx r_{\rm s}$ form
a clear feature, indicated by dashed curves in Figs.\
\ref{fig:f6}(b,c), with a pronounced blue horn (at 4-5.5 keV,
depending on $r_{\rm s}$) and a strongly suppressed red horn.  Note
also that the overall profiles always have  a strong blue peak around
6.5 keV, resulting from emission from larger ($\ga 7 R_{\rm g}$)
distances.  While energy of the  latter allows to constrain the
inclination angle, the energies and heights of the horns from $r_{\rm
d} \approx r_{\rm s}$ allow to study effects of space-time
metric. Namely, equation (\ref{einf}) implies that the gravitational
redshift of photons emitted from $r_{\rm d} \la 4$ depends on $a$
noticeably; e.g.\ for $\mu_{\rm obs} = 0.85$ and $r_{\rm d} = 4$,
photons are redshifted by $E_{\rm inf}/E_{\rm LN}=0.71$ for $a=0$,
while for $a=0.998$ their redshifts are in the range  $E_{\rm
inf}/E_{\rm LN}=0.66-0.77$ (for $a=0.998$ the redshift  depends on
photon initial direction via $p_{(\phi)}$). Then, $\approx 10$ per
cent difference in energies of the horns (formed at $r_{\rm d} \approx
r_{\rm s}$)  may be expected between low and high $a$. Furthermore,
for photons originating from $r_{\rm d} \la 4$, the space-time metric
affects trajectories, which is reflected in heights of the horns.

Indeed, such differences between model $S$ with $a=0.998$ and model
$S_0$, both with    $r_{\rm s}=4$, are clear in Fig.\
\ref{fig:f6}(c,f).   Note that these differences result only from GR
effects - in both cases the disc under the source has $v_{\rm K}
\simeq 0.5c$.    In Fig.\ \ref{fig:f6}(d,e) we show spectra for model
$S$ with $a=0$ (and the usual assumption of free fall within $r_{\rm
ms}$).  Due to additional Doppler redshift  ($v^r<0$)  the redshifted
horns have lower energies and are suppressed with respect to the blue
peak more significantly than in models $S_0$ with the same $r_{\rm s}$.

\begin{figure}
\centerline{\includegraphics[height=140mm]{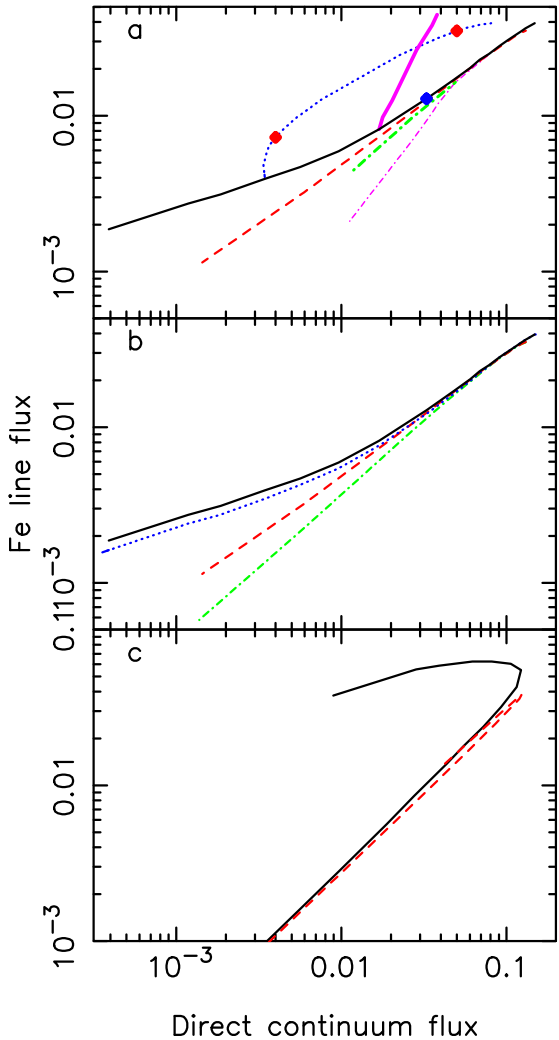}}
\caption{Total flux in the Fe K$\alpha$ line vs the primary emission
flux  observed at $\mu_{\rm obs}=0.85$. All models in panels (a) and
(b) assume a constant intrinsic luminosity of the primary source,
changes of the observed fluxes result from varying position of the
source (in general increasing distance corresponds to increase of the
fluxes).  {\bf (a)} The solid (black) curve is for model $S_{\rm K}$
with $a=0.998$ and the radial  distance between $r_{\rm s} = 1.5$ and
$r_{\rm s} = 100$; the (blue) diamond is for  $r_{\rm s} = 4$.  The
dashed (red) curve is for model $A$ with $a=0.998$, $V=0$ and $r_{\rm
s} = 1.6$--20.  The dotted (blue) curve is for model $C$ with $h_{\rm
s} = 0.07$--20; the (red) diamonds are for $h_{\rm s} = 1$ (left)  and
$h_{\rm s} = 8$ (right).  The heavier dot-dashed (green) curve is for
model $S_0$ and the thinner dot-dashed (magenta) curve is for model
$S_{\rm K}$ with $a=0$ (and a free fall inside $r_{\rm ms}$); both
models with $a=0$ are for  $r_{\rm s} = 3$--100 and assume $V=0.5$ at
$r_{\rm s}<6$. The heavy solid (magenta) curve is for constant
$\rho_{\rm s} = 3$, constant (Keplerian) angular momentum and $h_{\rm
s} = 0.07$--20.  All models in panel (a) take into account the
returning radiation.  {\bf (b)} The solid (black) and dashed (red)
curve is for model $S_{\rm K}$ and $A$, respectively, same as in panel
(a). The dotted and dot-dashed curves are for the same parameters  but
with neglected effect of returning radiation.  {\bf (c)}  The solid
(black)  and dashed (red) curve is for model $S_{\rm K}^{\rm PT}$
(i.e. for intrinsic luminosity changing with $r_{\rm s}$ according to
radial emissivity of a Keplerian disc) with $a=0.998$ and $a=0$,
respectively.  The primary source changes distance between $r_{\rm s}
= 1.5$ (for $a=0.998$) or 6 (for $a=0$) and $r_{\rm s} =100$ (toward
the bottom).  Normalisations of curves in (c) are chosen to yield the
same maximum values of the direct continuum flux.  }
\label{fig:f8}
\end{figure}

\begin{figure}
\centerline{\includegraphics[height=55mm]{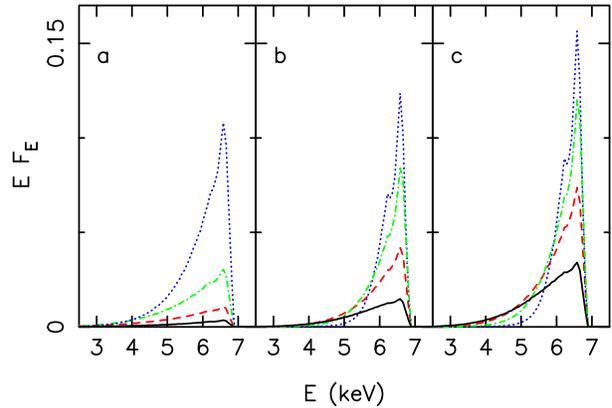}}
\caption{Changes of Fe K$\alpha$ line profile, for $\mu_{\rm obs} =
  0.85$, in models $A$ and $C$ with $a=0.998$.  {\bf (a)} Model $A$,
  $V=0$, with $r_{\rm s}=1.6$  (solid, black), 2 (dashed, red),  3
  (dot-dashed, green) and 6 (dotted, blue).  {\bf (b)} and {\bf (c)}
  are for model $C$ and $\hat C$, respectively, with $h_{\rm s}=2$
  (solid, black), 4 (dashed, red), 8 (dot-dashed, green) and 20
  (dotted, blue).  Profiles in panels (b) and (c) are rescaled by a
  factor 0.4.  }
\label{fig:f9}
\end{figure}

\subsection{The GR models of spectral variability}
\label{sec:var}

Uncorrelated variability between direct continuum and reflection
components (including the iron line) revealed in several objects is
difficult to understand if we observe  the same  primary emission that
illuminates the disc.  As proposed by Miniutti et al.\ (2003) and
Miniutti \& Fabian (2004), variability of both the observed continuum
and line may be induced by light bending if the primary source   is
concentrated near the axis of a Kerr black hole.  Moreover, they find
that variations of both components are not correlated  for some range
of parameters.

\begin{figure}
\centerline{\includegraphics[height=42mm]{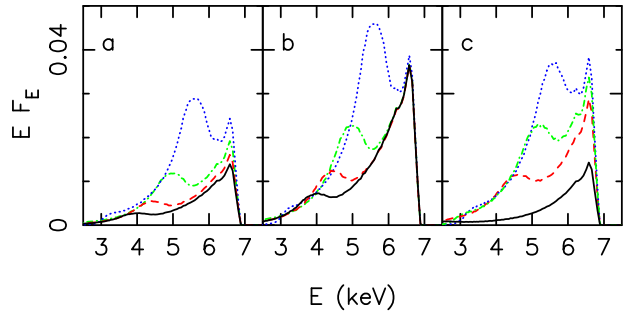}}
\caption{Changes of Fe K$\alpha$ line profile, for $\mu_{\rm obs} =
  0.85$, in model $S$.  Panels {\bf (a)} and {\bf (b)}  are for model
  $S_{\rm K}$ and $\hat S_{\rm K}$ (neglecting the time dilation),
  respectively, with $a=0.998$ and $r_{\rm s}=2.2$  (solid, black),
  2.5 (dashed, red),  3 (dot-dashed, green) and 4 (dotted,
  blue). Panel {\bf (c)} is for model $\hat S_0$ with $a=0$ and
  $r_{\rm s}=3$ (solid, black), 4 (dashed, red), 4.5  (dot-dashed,
  green) and 5 (dotted, blue).}
\label{fig:f10}
\end{figure}

Following Miniutti \& Fabian (2004), we compute the observed flux of
the primary source vs. that of the Fe  line, as the source position is
changed, for a source with constant intrinsic luminosity, see Fig.\
\ref{fig:f8}(a).  Changes of $\epsilon_{\rm Fe}$, related to varying
position of the source, in our basic models are shown  in
Fig.~\ref{fig:f4}; the corresponding changes of line profiles, for
model $A$, $C$ and $S_{\rm K}$, are shown   in Figs.~\ref{fig:f9}(a),
\ref{fig:f9}(b) and \ref{fig:f10}(a),  respectively.

\subsubsection{Model $A$}
\label{sec:var:a}

In model $A$, the primary continuum flux changes by two orders of
 magnitude when the distance of the source varies within  $r_{\rm
 s}=1.6-20$.  However, the line flux varies similarly in this model.
 At small $r_{\rm s}$, directly observed primary flux is significantly
 reduced while the illuminating flux enhanced by the light bending.
 The enhanced illumination of the disc is, however, concentrated in
 the innermost region (at $r_{\rm d} \la h_{\rm s}$, see Fig.\
 \ref{fig:f1})  and the reflected emission is subject to similar
 bending (due to which most of this emission is lost into black hole)
 as the primary emission.  As a result, the line observed at low
 inclinations [Fig.\ \ref{fig:f9}(a)] roughly follows changes of the
 continuum. Variations of  the line flux are reduced by a factor of
 $\approx 2$ with respect to the continuum  flux, primarily due to
 returning radiation, see Fig.\ \ref{fig:f8}(b).

\subsubsection{Models $S$}
\label{sec:var:s}

The most significant reduction of variability of the line flux seems
to characterise model $S_{\rm K}$, with $a=0.998$ and $r_{\rm s}<4$,
where the line flux changes by a factor of 7 while  the primary
continuum flux changes by two orders of magnitude.  On the other hand,
in model $S_0$ (with the same kinematic properties but $a=0$) both
components are linearly correlated, while in model $S_{\rm K}$ with
$a=0$ (involving free fall within $r_{\rm ms}$) inward acceleration in
the plunging region results in increasing decline of the line flux;
see the dot-dashed curves in Fig.~\ref{fig:f8}(a).  For $r_{\rm s}>6$
model $S_{\rm K}$ yields the same relation between the line and
continuum fluxes (regardless of $a$)  as model $A$ with $V=0$.

We stress that the reduced variability of the line flux, for small
$r_{\rm s}$ and high $a$, is directly related to properties of the
Kerr metric. Namely,  decreasing $r_{\rm s}$ results in stronger
bending to equatorial plane and this in turn gives rise to enhanced
illumination of the disc surface at $r_{\rm d} > 6$, where photons
forming the blue peak around 6.5 keV come from, see Fig.\
\ref{fig:f6}(b,c).  As a result the line profiles, Fig.\
\ref{fig:f10}(a), may be decomposed into two components: (i) the
redshifted emission from the region under the source, which changes
similarly to the primary emission; and (ii) the blue peak  with much
lower variability.

\begin{figure}
\centerline{\includegraphics[height=50mm]{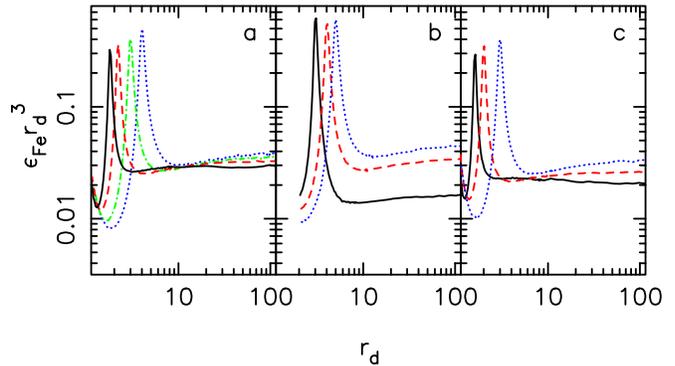}}
\caption{Changes of $\epsilon_{\rm Fe}(r)$ resulting from the change
of distance of primary source in models $\hat S$ (with neglected time
dilation).  Panels (a) and (c) show impact of azimuthal velocity
pattern, Keplerian vs constant with $V=0.5$, for $a=0.998$.  Panels
(b) and (c) show effect of black hole spin, $a=0$ vs 0.998, for
constant $V=0.5$.  {\bf (a)}   The solid (black), dashed (red),
dot-dashed (green) and dotted (blue) curves are for model $\hat S_{\rm
K}$  with $a=0.998$ and $r_{\rm s}=1.8$, 2.2, 3 and 4.  {\bf (b)} The
solid (black), dashed (red) and dotted (blue) curves are for model
$\hat S_0$ with $V=0.5$, and $r_{\rm s}=3$, 4 and 5.  {\bf (c)} The
solid (black), dashed (red) and dotted (blue) curves are for model
$\hat S$ with $a=0.998$, $V=0.5$ and $r_{\rm s}=1.6$, 2 and 3.  }
\label{fig:f11}
\end{figure}

The distinct properties of the Kerr metric, underlying reduction of
the line variability in model $S$,  are clearly illustrated in models
neglecting the effect of  time dilation at the primary source, i.e.\
with the source luminosity $= L_0/g_t$, where $g_t$ is given by
equation (\ref{dtdt}); note that the time dilation  affects in the
same way both the directly observed  and illuminating flux, thus
increasing magnitude of variability of both the primary and reflected
components, but does not change the relative variability  of these two
components.

The $\epsilon_{\rm Fe}(r)$ profiles in models neglecting time dilation
are shown in Fig.~\ref{fig:f11}. In model $\hat S_{\rm K}$ with
$a=0.998$, focusing of primary emission toward the disc yields
constant $\epsilon_{\rm Fe}(r)$ in the region where photons forming
the blue peak originate. This, in turn, results in identical line
profiles at $E>6$ keV (Fig.~\ref{fig:f10}(b)), while the primary
emission flux changes (similarly to the line flux at $<6$ keV).  A
similar neglect of time dilation in model $\hat S_0$ does not  result
in any range of $r_{\rm d}$ with similarly constant $\epsilon_{\rm
Fe}$, Fig.~\ref{fig:f11}(b); in this case flux in the blue peak is
strictly correlated with the primary flux, see Fig.~\ref{fig:f10}(c).

Note also that $V$ increases with decreasing $r_{\rm s}$ in model
$S_{\rm K}$, which seems to be the most likely  for a source  close to
the disc surface, as it may be expected to be rigidly coupled to the
accretion disc by magnetic fields.  In general, the observed line and
continuum fluxes are sensitive to assumed $V$, see Section 3.3 and
discussion of model $C$ below. We emphasise, however, that the
increase of $V$ in model $S_{\rm K}$ is not crucial for reduction of
variability of the line.    For high $a$ and small $r_{\rm s}$,
bending to the disc plane is a robust effect, exceeding in magnitude
other relevant effects. In particular, approximate constancy of
$\epsilon_{\rm Fe}(r)$ around $r_{\rm d}=10$ is achieved also for
fixed $V$,  see Fig.~\ref{fig:f11}(c).

Our simplified description of generation of X-rays by a point source
with constant intrinsic luminosity allows to illustrate various GR and
SR effects, as in panels (a) and (b) of Fig.\ \ref{fig:f8}.
Combination of these effects with more realistic modelling of the
X-ray source may result in variety of additional properties. A
particularly interesting case of  model $S_{\rm K}^{\rm PT}$, with
intrinsic luminosity proportional to the energy release rate in a
Keplerian disc, is shown in Fig.~\ref{fig:f8}(c).  Effects due to
bending to the equatorial plane are again clear in comparison of
models involving $a=0$ and $a=0.998$. In the former case, both the
line and continuum fluxes increase with decreasing $r_{\rm s}$ ($>10$)
due to increase of the dissipation rate, however for $r_{\rm s} < 10$
the fluxes decrease, due to decrease of both the dissipation rate and
photon escape probability, following approximately the same relation
between the continuum and line flux as at higher $r_{\rm s}$.  On the
other hand, for $a=0.998$ the primary emission flux achieves the
maximum value around $r_{\rm s} = 3$ and than decreases by a factor of
3 with the distance decreasing to $r_{\rm s}=1.8$,  while the Fe flux
remains almost constant.

\begin{figure}
\centerline{\includegraphics[height=45mm]{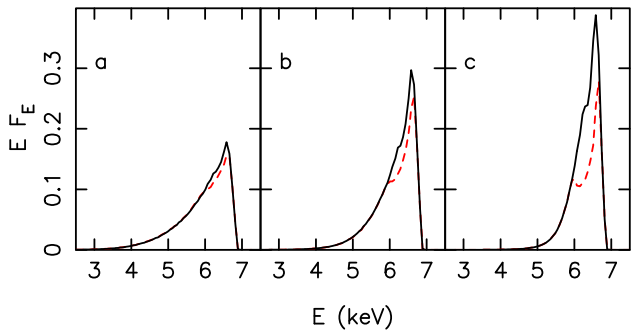}}
\caption{The Fe K$\alpha$ line profiles, for $\mu_{\rm obs} = 0.85$,
  in model $C$ with  $h_{\rm s}=4$ (a), 8 (b)  and 20 (c).  The solid
  (black) profiles are for $r_{\rm out} = 600$ and the dashed (red)
  are for $r_{\rm out} = 100$.  }
\label{fig:f12}
\end{figure}

\subsubsection{Model $C$}
\label{sec:var:c}

Miniutti et al.\ (2003) and  Miniutti \& Fabian (2004) show that with
modifications of the model with source close to the axis (see
definition of our model $C$),  variability of the line flux is
strongly reduced over some range of heights. Specifically,  they
identify three different regimes in which the  iron line is correlated
(at small $h_{\rm s}$),  almost constant (at intermediate $h_{\rm s}$)
and anti-correlated (at high $h_{\rm s}$) with respect to the direct
continuum.  For $\mu_{\rm obs}=0.85$ and  $\rho_{\rm s}=2$, these
regimes correspond approximately to $h_{\rm s} = 1-4$,  $h_{\rm s} =
4-13$ and $h_{\rm s} >13$, respectively.
 
The dotted curve in Fig.\ \ref{fig:f8}(a) shows the iron line flux
[changes of the line profile are shown in Fig.\ \ref{fig:f9}(b)] as a
function of primary continuum flux for model $C$.  In agreement with
Miniutti \& Fabian (2004) we find that the primary continuum changes
by a factor of 20 for the height varying between $h_{\rm s}=1$ and 20.
On the other hand, we find different dependence of the line flux on
$h_{\rm s}$, in particular we do not note a decrease of the line flux
at large heights.  We suspect that this discrepancy results primarily
from different  values of the outer radius of the disc, $r_{\rm out} =
100$ assumed by  Miniutti \& Fabian (2004) and $r_{\rm out} = 600$ in
this paper. An additional, small difference involves  angular law of
Fe emission (assumed to be locally isotropic in Miniutti \& Fabian
2004), which yields a moderate ($\la 20$ per cent, see Fig.\
\ref{fig:f18}) change of quantitative results.  On the other hand,
value of $r_{\rm out}$ appears to be crucial for qualitative
predictions of the model.

The finite extension of the outer radius of the disc results in the
presence of a central dip in the line profile, which is not important
for the total flux of the line for sufficiently large  $r_{\rm
out}$. However, at $h_{\rm s}>10$, the solid angle subtended by a disc
with $r_{\rm out} = 100$ is significantly smaller than, e.g.\ for
$r_{\rm out}=1000$ (see Matt et al.\ 1992). Moreover, illumination is
shifted to outer regions of the disc in model $C$ due to azimuthal
motion. As a result, the line flux is increasingly underestimated with
increasing $h_{\rm s}$, e.g.\  by  over 30 per cent for $h_{\rm
s}=20$, in the model with $r_{\rm out} = 100$, see Fig.\
\ref{fig:f12}.  We conclude that for face-on observer the
anti-correlation is an artificial effect which does not occur if
sufficiently large $r_{\rm out}$ is taken into account.  On the other
hand, we find  a real anti-correlation  for higher inclinations, see
below.

Obviously, illumination of distant regions, and hence the choice of
specific value of $r_{\rm out}$, becomes more important for increasing
$h_{\rm s}$.  Our aim is to study effects related to  GR or SR rather
than simple geometric effects related to  the change of the solid
angle. Therefore, we assume $r_{\rm out} = 600$, for which  reflection
from $r_{\rm d}>r_{\rm out}$ is negligible (for $h_{\rm s} \le 20$),
in our computations. Clearly, such a value is necessary for
applications  of the model to stellar-mass black hole systems (cf.\
Miniutti, Fabian \& Miller 2004).  For large black hole masses, $M >
10^7 M_{\sun}$, part of reflection from distant regions of the disc
would be subject to significant delays,  owing to light travel-time
effects, see Section 2.  In such cases,  variations of the reprocessed
radiation should be somewhat reduced due to mixing of radiation
responding to primary emission from various positions of the
source. On the other hand,  for large $h_{\rm s}$, the Fe flux is
close to constant even without the mixing effect. A more interesting
effect, related to light-travel time  in model $C$, for high values of
$M$, would involve  response of the line profile to changes of the
source position.  In particular, rapid increase of $h_{\rm s}$ should
result in such changes   of the line profile  as those predicted by
Stella (1990). For black holes with lower masses, $M \la 10^7
M_{\sun}$, impact of time delays on variability properties depends
on the range of heights and the time-scale over which the variability
is analysed. Note, e.g., that for $M=6 \times 10^6 M_{\sun}$ (which is
within the range of masses allowed in MCG--6-30-15 by optical
measurements, see  Section \ref{sec:mcg}), the mixing effect could not
be neglected  in applications of model $C$ with $h_{\rm s} \ge 10$.

For $r_{\rm out}=600$, taking into account the whole range of heights
($\le 20 R_{\rm g}$), we find the following relation between the line
and continuum in model $C$.  At  $h_{\rm s}<1$ the line varies by a
factor of 2 while the primary continuum is almost constant.  For
$h_{\rm s} =1$--8, variations of the line are reduced by a factor of 2
with respect to the continuum. For $h_{\rm s}>8$, the line flux is
close to constant.  In the last case, majority of Fe photons come from
the region of disc where light bending and beaming effects are small
and  changes of line flux result from varying time dilation at the
primary source. Neglecting the time dilation we find that  the line
flux  is constant  while the primary flux increases by 50 per cent
with  increase of height  from $h_{\rm s}=8$ to $h_{\rm s}=20$. Note,
however, that although the total line flux does not change, the line
profiles change significantly with $h_{\rm s}$ [see Fig.\
\ref{fig:f9}(c)] -  in general, flux at 5.8 keV remains approximately
constant but the wing below this energy  weakens while the blue peak
strengthens with the increase of $h_{\rm s}$ [note the change of the
distance where most of Fe photons come from in Fig.\ \ref{fig:f4}(b)].

We emphasise that the above behaviour is not a generic prediction,
resulting from light bending,  for a model involving a source with
changing $h_{\rm s}$. It is strongly related to specific kinematic
assumptions underlying model $C$, see Section 2.2.3.  Apart from
impact on the line flux, discussed in Section 3.3, changes of $V$
reduce  variability of the primary continuum at small heights (where
weaker light bending is balanced by stronger SR beaming with
increasing $h_{\rm s}$) and enhance its variability at $h_{\rm s}>2$.
Assuming another pattern of motion, we obtain different  relations
between the spectral components. E.g., for a source changing $h_{\rm
s}$ at constant $\rho_{\rm s}$,   but with constant angular momentum
equal to angular momentum on a Keplerian orbit in the equatorial plane
at $\rho_{\rm s}$, $V$ increases monotonically with $h_{\rm s}$. In
this case, the magnitude of  the line flux variability is higher than
that of the continuum at all $h_{\rm s}$, see the heavy solid curve in
Fig.\ \ref{fig:f8}(a). On the other hand,  setting $V=0$ in model $C$
we find, for  $h_{\rm s}>1$, the same relation between the line and
primary  flux  as in model $A$, i.e. shown by dashed curve in Fig.\
\ref{fig:f8}(a).

\begin{figure}
\centerline{\includegraphics[height=72mm]{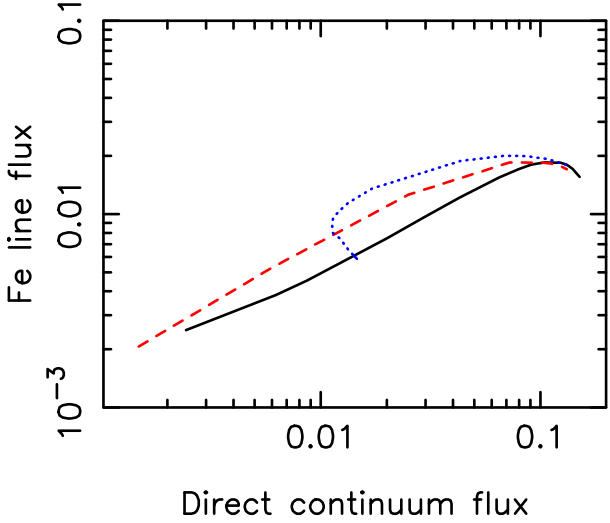}}
\caption{Same as Fig.\ \ref{fig:f8}(a) but  for an observer
inclination $\mu_{\rm obs}=0.45$. Solid (black) curve is for model
$S_{\rm K}$, dotted (blue) for model $C$, dashed (red) for model $A$.
All models assume $a=0.998$.  }
\label{fig:f13}
\end{figure}

\subsubsection{Summary of variability models}

In summary, increased illumination of the innermost disc, within a few
$R_{\rm g}$, resulting from deflection of light {\it toward the
centre}, is not sufficient to yield reduced variability of reflection
component for small inclination angles. The major constraint results
from the fact that reflection from this innermost region is not
observed at small inclinations.  Then, reduction of observed
variability requires enhanced illumination of disc beyond a few
$R_{\rm g}$, which can be achieved  in the Kerr metric for small
$r_{\rm s}$ as a result of bending {\it toward the equatorial plane}.
Interestingly, the same mechanism reduces variations of the line for a
source at small heights at the axis, through the effect of returning
radiation, as this radiation illuminates the disc at distances larger
than the source height.  Alternatively, a more extended  illumination
may result from the SR beaming and then the reduced variability can be
achieved for appropriate pattern of motion, e.g.\ for $V$  decreasing
with increasing $h_{\rm s}$.

On the other hand, the above constraint does not apply to systems with
higher inclinations ($\cos \theta_{\rm obs} < 0.6$) and for such
systems  the line flux  varies with much smaller amplitude than the
continuum flux in all models, see Fig.~\ref{fig:f13}.  Moreover,
anticorrelation between the line and continuum fluxes observed at such
$\theta_{\rm obs}$ is a generic property of all models at large
$r_{\rm s}$ or $h_{\rm s}$.  The decrease  of the line flux, with
increasing $r_{\rm s}$, is a purely SR effect,  resulting from (i)
angular distributions of the line flux, $F(\mu_{\rm em})$, in the disc
rest frame,  which declines with decreasing $\mu_{\rm em}$; (ii)
increase of the distance, $r_{\rm d}$, where     most Fe photons come
from with increasing $r_{\rm s}$; and (iii) decrease of the orbital
velocity  of disc, with increasing $r_{\rm d}$, and thus less
efficient  beaming to high $\theta_{\rm obs}$.

\begin{figure}
\centerline{\includegraphics[height=56mm]{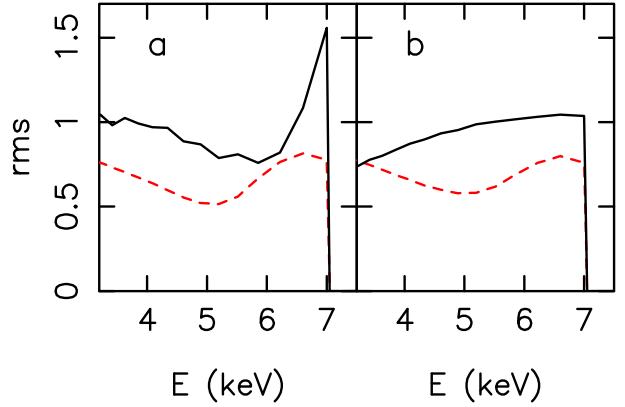}}
\caption{The rms spectra for Fe K$\alpha$ line observed at $\mu_{\rm
obs}=0.85$.  {\bf (a)} Solid (black) curve is for model $S_{\rm K}$
with the radial distance changing between $r_{\rm s}=1.5$ and $r_{\rm
s}=100$.  Dashed (red) curve is for model $C$ with $h_{\rm
s}=0.07$--20.  {\bf (b)}  Solid (black) and dashed (red) curve is for
model $A$ with $V=0$ and $h_{\rm s}=1.6$--3 and $h_{\rm s}=1.6$--20,
respectively. All models assume $a=0.998$.}
\label{fig:f14}
\end{figure}

\begin{figure}
\centerline{\includegraphics[height=104mm]{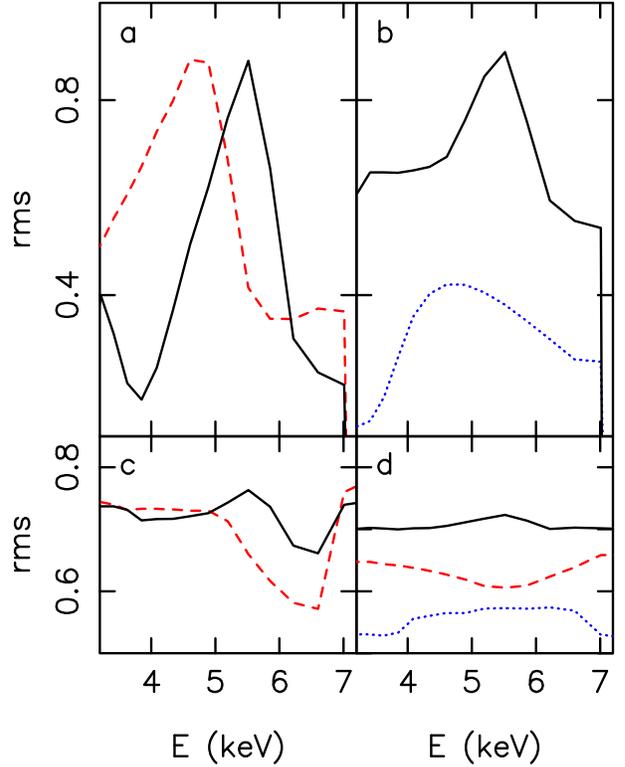}}
\caption{The rms spectra for Fe K$\alpha$ line (top panels) and
superposition of Fe line and primary continuum (bottom panels)
observed at $\mu_{\rm obs}=0.85$.  {\bf (a,c)} The dashed (red) and
solid (black) curves are for model $S_{\rm K}$, $a=0.998$,  with
$r_{\rm s}=1.5$--3 and $r_{\rm s}=2.2$--4, respectively.  {\bf (b,d)}
The dotted (blue)  curves are for model $C$ with $a=0.998$ and $h_{\rm
s}=0.07$--1.   The solid (black) curves are for model $S$ with $a=0$
and $r_{\rm s}=3$--5.    The dashed (red) curve in panel (d) is for
model $C$ with $h_{\rm s}=0.07$--20 (i.e.\ the same parameters as for
the dashed curve in Fig.~\ref{fig:f14}(a)). Some spectra in panels (c)
and (d) are shifted vertically for clarity.}
\label{fig:f15}
\end{figure}
\subsection{The fractional variability}
\label{sec:rms}

We point out that model $S$ with high $a$ is uniquely capable to
explain  suppressed variability around 6.5 keV  in energy-dependent
patterns of spectral variability.  We focus here on effects related to
changes of the Fe line profile.  A more detailed analysis of spectral
variability, including the Compton reflection component, will be
presented in our next paper.

The fractional variability of observed spectra,  as a function of
energy, is quantitatively characterised by the root mean square (rms)
variability function. In models relating variability to change of
position  of the primary source, we determine the rms spectra
according to the following definition
\begin{equation}
{\rm rms}(E) =  { 1\over <f(E)>} {\sqrt
{\sum_{i=1}^{N}{[f_i(E)-<f(E)>]^2 \over N-1}} },
\end{equation}
where the summation is performed over $N$ positions of the primary
source, $f_i(E)$ is the  photon flux in the energy band, $E$,
corresponding to $i$-th position of the source and  $<f(E)>$ is the
average photon flux in this band for all $N$ positions.

In Figs.~\ref{fig:f14} and \ref{fig:f15}(a,b) we show the rms spectra
for the line  and in Figs.\ \ref{fig:f15}(c,d) for superposition of
the line and the primary power-law.   Taking into account the full
ranges of source positions considered in our basic models - i.e.\
radial distances $r_{\rm s}=1.5$--100 and $r_{\rm s}=1.6$--20 in
models $S$ and $A$, respectively, and the height $h_{\rm s}=0.07$--20
in model $C$ - we find that the rms spectra for the line increase
toward low energies below 5 keV, see Fig.~\ref{fig:f14}, which effect
is related to varying strength and extent of the red wing of the
line. However, the line flux at $<5$ keV is typically much  lower than
the primary continuum flux, therefore these variations of the red wing
do not produce noticeable excesses in the rms spectra of the line plus
power-law.  On the other hand, both model $A$ and $C$ are
characterised by reduced variability of the line flux around 5.5 keV
(see Fig.\ \ref{fig:f9}(b,c)) which gives rise to declines in the rms
spectra both for the line  (Fig.~\ref{fig:f14}) and for the line +
power-law (dashed curve in Fig.\ \ref{fig:f15}(d)).  In turn, in model
$S_{\rm K}$ at $r_{\rm s} > 10$, the line is formed in majority by
photons from  $r_{\rm d} \approx r_{\rm s}$ and the energy of the blue
peak changes with $r_{\rm s}$, giving rise to strong increase of rms
spectrum at $E>6$ keV, see Fig.~\ref{fig:f14}(a).

In most models, even these predicting the reduced variability  of the
total line flux, a decline around 6.5 keV  does not occur in the rms
spectrum. We notice such suppression only in model $S_{\rm K}$  (and
similarly $S_{\rm K}^{\rm PT}$) with $a=0.998$ and $r_{\rm s} < 4$,
where bending to the disc plane yields strongly reduced variability of
the blue peak, see discussion in Section \ref{sec:var:s}.  The
detailed shape of resulting signal in the rms spectrum depends on the
assumed range of $r_{\rm s}$, see Fig.~\ref{fig:f15}(a,c). The range
of distances extending deeper into the ergosphere corresponds to
stronger suppression around 6.5 keV. For $r_{\rm s}=2$--4,  decline
above 6 keV is less pronounced, on the other hand, a strongly variable
and relatively strong red wing of the line  gives rise to an excess
variability around 5.5 keV, see the solid curves in
Fig.~\ref{fig:f15}(a,c).

Direct relation of these effects with properties of the Kerr metric is
again  clearly illustrated by comparing models $S_0$ and $S_{\rm K}$
with $a=0.998$.  The rms spectra for model $S_0$ with  a (rather
unphysical) assumption of the primary source changing the radial
distance at $r_{\rm s}<6$ are shown by the solid curves in
Figs.~\ref{fig:f15}(b,d).  For $a=0$, variable red wing gives rise to
excess at 5.5 keV, however,  the blue peak  changes similarly to the
primary continuum and the rms spectrum is flat above 6 keV.

\begin{figure}
\centerline{\includegraphics[height=60mm]{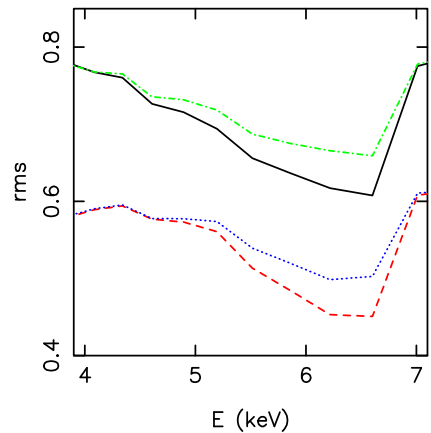}}
\caption{The rms spectra for the superposition of Fe line and primary
continuum observed at $\mu_{\rm obs}=0.85$, in model $S_{\rm K}^{\rm
PT}$. The upper and lower pair of curves is for  $r_{\rm s}=1.7-2.2$
and $r_{\rm s}=1.8-3$, respectively.  The solid (black) and dashed
(red) curves are for  $I(\mu_{\rm em}) \propto 1 + 2.06 \mu_{\rm em}$.
The  dot-dashed (green) and dotted (blue) curves are for  $I(\mu_{\rm
em}) \propto \log (1 + 1/ \mu_{\rm em})$.  }
\label{fig:f19}
\end{figure}

We did not find significant differences between predictions of models
$S_{\rm K}$ and $S_{\rm K}^{\rm PT}$, see Fig.\ \ref{fig:f19}. On the
other hand, the quantitative properties are slightly dependent on the
angular emissivity law, see next section.

Finally, note that source located close to the disc surface (as
required to model a pronounced red wing, see Section \ref{sec:mcg}),
but changing the height rather than the radial distance, gives rise to
spectral variability  significantly different than model $S_{\rm
K}$. In particular, in model $C$  with small  $h_{\rm s}$, the line
varies much more significantly than the primary emission [see dotted
curve in Fig.~\ref{fig:f8}(a)] yielding an extended (over $\Delta E
\approx$ 3 keV)   excess in the rms spectrum, see dotted curves in
Fig.~\ref{fig:f15}(b,d).  In model $A$ with small $r_{\rm s}$ (see
Figs.~\ref{fig:f9}(a) and \ref{fig:f14}(b)) the most significant
changes of the line occur close to the blue peak.

\begin{figure}
\centerline{\includegraphics[height=46mm]{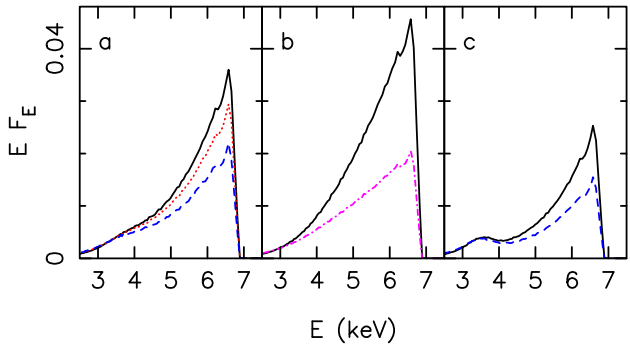}}
\caption{Fe K$\alpha$ line profiles observed at $\mu_{\rm obs} = 0.85$
for various  angular laws.  In all panels the solid (black) profiles
are for limb darkening in electron scattering limit, $I(\mu_{\rm em})
\propto 1 + 2.06 \mu_{\rm em}$. {\bf (a)} Model $C$ with $h_{\rm
s}=0.4$. The dotted (red) curve is for locally isotropic emission,
$I(\mu_{\rm em})=I_0$;  the dashed (blue) curve is for $I(\mu_{\rm
em}) \propto \log (1 + 1/ \mu_{\rm em})$.  {\bf (b)} Model $C$ with
$h_{\rm s}=1$. The dot-dashed (magenta) profile is for optically thin
emission, $I(\mu_{\rm em}) \propto 1 / \mu_{\rm em}$. {\bf  (c)} Model
$S$ with $r_{\rm s}=2$ and $V=0.5$;  the dashed (blue) curve is for
$I(\mu_{\rm em}) \propto \log (1 + 1/ \mu_{\rm em})$.  }
\label{fig:f16}
\end{figure}

\subsection{Limb darkening}
\label{sec:limb}

As noted by Beckwith \& Done (2004), for a fixed  radial emissivity
change of the limb darkening may strongly influence the line profile.
Therefore, we have checked our results for various angular laws
considered in literature.

The limb darkening in the form   $I(\mu_{\rm em}) \propto 1 + 2.06
\mu_{\rm em}$, assumed in most of our calculations,  approximates
emission from a disc with ionised atmosphere.   The same angular
distribution is assumed  in the {\sc laor} model (see Laor 1991),
which is often applied in spectral analysis.    Emission from a
uniform, not ionised slab is approximated by $I(\mu_{\rm em}) \propto
\log(1+1/\mu_{\rm em})$, Ghisellini, Haardt \& Matt (1994).  Emission
from an optically thin material has $I(\mu_{\rm em}) \propto 1 /
\mu_{\rm em}$.  Finally, a locally isotropic emission, $I(\mu_{\rm
em})=I_0$, is sometimes considered for simplicity.
 
Fig.\ \ref{fig:f16} shows line profiles for various angular emission
laws.  Note that change of the angular law results, in some cases, in
a significant change of  height of the blue peak, while the lowest
energy ($\la 3.5$ keV) part is roughly unaffected. Then, the slope of
the red wing - connecting the low  energy part with the blue peak -
changes, following the change of the height of the latter.  The
following effects result in such properties of the line.

For $\mu_{\rm obs}=0.85$, the change of the angular law turns out to
be unimportant for   the magnitude of contribution of photons emitted
from the disc within the ergosphere (forming the observed line  at
$E_{\rm inf} \la 3.5$ keV). In this part of the disc,  only photons
emitted  into a narrow range of initial  directions, with intermediate
values  of $\mu_{\rm em}=0.4-0.6$, are observed at $\mu_{\rm
obs}=0.85$.  Then, flux emitted in these directions  remains roughly
unaffected by the change of angular law, under the condition that
total energy of fluorescent  photons emitted from unit area of the
disc remains constant.

On the other hand, the blue peak is mostly due to emission from outer
regions, where  light bending is not crucial.  Then, changes in the
height of the peak simply correspond to change of the flux emitted to
$\mu_{\rm em} \approx \mu_{\rm obs}$, again under the condition  that
the total emitted energy remains constant.  E.g.,  $I(\mu_{\rm em})
\propto 1 + 2.06 \mu_{\rm em}$ yields $F(\mu_{\rm em}=0.85)$ higher by
a factor of $\approx 2$, 1.7 and 1.2  than $I(\mu_{\rm em}) \propto
1/\mu_{\rm em}$, $I(\mu_{\rm em}) \propto \log(1+1/\mu_{\rm em})$ and
$I(\mu_{\rm em})=I_0$, respectively. These factors are in a rough
agreement with differences in the heights of the blue peaks  in Fig.\
\ref{fig:f16}.

\begin{figure}
\centerline{\includegraphics[height=50mm]{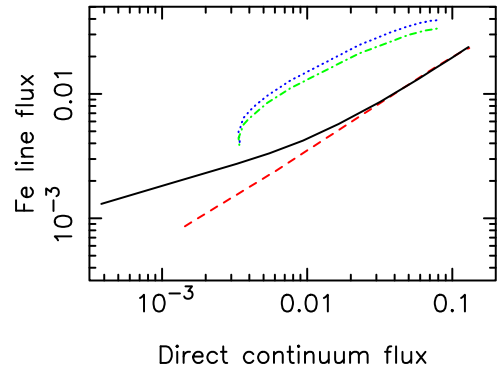}}
\caption{Total flux in the Fe K$\alpha$ line vs the primary emission
flux observed at $\mu_{\rm obs}=0.85$. The solid (black) and dashed
(red) curve is for model $S_{\rm K}$ and $A$, respectively, with 
$a=0.998$ and constant intrinsic luminosity, same as in Fig.\ 
\ref{fig:f8}(a), but with 
$I(\mu_{\rm em}) \propto \log (1 + 1/ \mu_{\rm em})$. The dotted
(blue) and dot-dashed (green) curve is for model $C$ with  $I(\mu_{\rm em})
\propto 1 + 2.06 \mu_{\rm em}$ (same as in Fig.\ \ref{fig:f8}(a)) and
$I(\mu_{\rm em})=I_0$, respectively. }
\label{fig:f18}
\end{figure}

In conclusion, we confirm that change of the angular law results in
the change of the line profile, mostly in the strength of the blue
peak.  However, we find that these (systematic) changes of the line do
not affect the main conclusions of this paper (assuming that the same
angular law characterises  emission from the whole disc surface).
Specifically, the qualitative properties of variability  of the total
flux of the line as well as the rms spectra,  are negligibly affected
by the choice of a particular law.  Figs.\ \ref{fig:f19} and
\ref{fig:f18} (in comparison with Fig.\ \ref{fig:f8})  illustrate
changes of such properties corresponding to the change of $I(\mu)$
from limb darkened to limb brightened. A slightly weaker signal in the
rms spectrum, for the latter, appears to be the most significant
effect resulting from such a change of    $I(\mu)$.  Note also that
changes of the line profile, at $E<5$ keV,  resulting from the change
of the angular law are much smaller than changes resulting from
significant change of $h_{\rm s}$ or $r_{\rm s}$, compare Figs.\
\ref{fig:f9} and \ref{fig:f16}, making our conclusions on modelling
the extended red wing in the next section independent  of  angular
emission law. Only assumption of $I(\mu_{\rm em}) \propto 1 / \mu_{\rm
em}$ (Fig.\ \ref{fig:f16}(b)) could affect our conclusions more
significantly, however, such an (optically thin) emission from an
accretion disc seems very unlikely.

\section{MCG--6-30-15}
\label{sec:mcg}

The profile of broad Fe line in the X-ray spectrum of Seyfert 1 galaxy
MCG--6-30-15, first clearly resolved by {\it ASCA} (Tanaka et al.\
1995), is consistent with that  predicted from an accretion disc
inclined at $30 \degr$. In part of the {\it ASCA} observation, when
the 2--10 keV flux was 2 times lower than average, $F_{2-10}=4 \times
10^{-11}$ erg/s/cm$^2$,  for the whole  observation, the line extended
below 4 keV (Iwasawa et al.\ 1996; that low flux state was defined as
the deep minimum). Profiles  of the Fe line, extending below 4 keV,
were revealed also in two {\it XMM-Newton} observations (Wilms et al.\
2001; Fabian et al.\ 2002).

During the first {\it XMM-Newton} observation, in 2000, the average
flux,  $F_{2-10}=2.4 \times 10^{-11}$ erg/s/cm$^2$, was close to that
of the deep minimum.  Then, Wilms et al.\ (2001) and Reynolds et al.\
(2004) argued that this observation  caught the source in its deep
minimum state. However, the source was highly variable, changing the
flux between $F_{2-10}= 1.5$ and $4.8 \times 10^{-11}$ erg/s/cm$^2$,
i.e. below the deep minimum and above the typical flux levels from
that source, respectively. During  the second, longer observation with
{\it XMM-Newton}, in 2001, with the average  $F_{2-10} = 4.1 \times
10^{-11}$ erg/s/cm$^2$, the flux again varied, by a factor of 5,
between values lower than that of the deep minimum and higher than the
typical average value. Using the {\it RXTE} observations Vaughan \&
Fabian (2004; see  their figure 1 and related discussion) argue that
in both cases the source was observed in its typical state.

We focus below on the second {\it XMM-Newton} observation.  A
phenomenological model with a broken power-law radial emissivity (see
below) gives an excellent fit to the data, with $\chi_{\nu}^2 < 0.9$
(Fabian et al. 2002), therefore, we presume that this model reproduces
properties of the emitting region with high precision and we take
parameters of the best fit as basic for our discussion.

The most tightly  constrained value of the black hole mass in
MCG--6-30-15, from X-ray variability, is  $\sim 3 \times 10^6
M_{\sun}$ (in agreement with optical measurements yielding  values
between 3 and $6 \times 10^6 M_{\sun}$; McHardy et al.\ 2005).
 
\subsection{Time-averaged profile}
\label{sec:mcg:profil}

\begin{figure}
\centerline{\includegraphics[height=76mm]{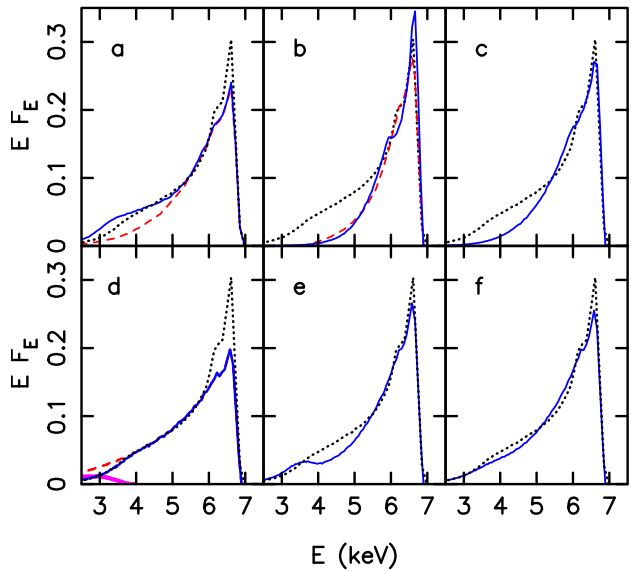}}
\caption{In all panels the dotted (black) curve shows profile of the
Fe line derived  with parameters given in Fabian et al.\ (2002), see
text. {\bf (a)} The solid (blue) curve is for parameters of the best
fit in Vaughan \& Fabian (2004), i.e.\ $q_{\rm in}=6.9$, $q_{\rm
out}=3$, $r_{\rm br}=3.4$, $r_{\rm in}=1.8$. The dashed (red) curve is
for a model with $q=3$, $r_{\rm in}=1.2$ and $a=0.998$.  The solid
(blue) curves in panels {\bf (b)} and {\bf (c)} are for model $C$ with
$h_{\rm s}=8$ and 4, respectively, $r_{\rm out}=100$ and $I(\mu_{\rm
em})=I_0$  (as in Miniutti et al.\ 2003). The dashed (red) curve in
{\bf (b)} is for model $A$ with $a=0$, $V=0$, $r_{\rm s}=3$ and
$r_{\rm in}=6$ ($= r_{\rm ms}$).  The dashed (red) curve in panel {\bf
(d)} is for model $A$ with $r_{\rm s}=1.6$, $a=0.998$ and
$V=0$. Contribution to this profile from the region of the disc within
and beyond $r_{\rm d}=2$ is shown by the heavy solid (magenta) and the
thinner solid (blue) curve, respectively.   {\bf (e)} The solid (blue)
curve is for model $S_{\rm K}$ with $r_{\rm s}=2$.  {\bf (f)} The
solid (blue) curve is for model $C$ with $h_{\rm s}=0.4$. All
profiles, except for solid in panels (b) and (c), correspond to
$I(\mu_{\rm em}) \propto 1 + 2.06 \mu_{\rm em}$. }
\label{fig:f7}
\end{figure}

As noted in Section \ref{sec:wing}, detection of photons with energies
$<4$ keV   is considered as the evidence of rapid rotation of a black
hole.  The best studied example of a relativistic Fe line - with
elongated wing formed by such strongly redshifted photons - comes from
{\it XMM-Newton} observation of MCG--6-30-15 (Fabian et al.\ 2002).
This is best fitted with a steep ($q_{\rm in}=4.8$) emissivity inside
$r_{\rm br}=6.5$ and  a flatter ($q_{\rm out}=2.5$) profile beyond
that, and with the inner radius of emitting region $r_{\rm in}=2$; the
fit was obtained with the   {\sc laor} model (in the X-ray data
fitting code {\sc xspec}),  which assumes  $r_{\rm out}=400$ and
$I(\mu_{\rm em})  \propto 1 + 2.06 \mu_{\rm em}$,  see Laor (1991).
The dotted profiles in all panels of Fig.\ \ref{fig:f7} correspond to
those best fitting parameters.

We note that there is a degeneracy between   $q_{\rm in}$ and $r_{\rm
br}$, in general higher $q_{\rm in}$ allows for lower $r_{\rm
br}$. Vaughan \& Fabian (2004) fit  the same {\it XMM} observation
(with the data set extended by including the EPIC pn data) with
$r_{\rm br}=3.4$ and $q_{\rm in}=6.9$, the solid curve in Fig.\
\ref{fig:f7}(a).  Nevertheless, both sets of parameters require the
line flux at 4 keV to be two times larger than in a model assuming a
single power-law radial emissivity with $q=3$ (see Fig.\
\ref{fig:f7}(a); the three curves are normalised  to yield the same
flux in 5-6 keV range). In summary, the data seem to require a steep
emissivity in the innermost region to explain  strength of the red
wing. At larger distances the emissivity has to flatten to produce a
rather strong blue peak.

Such steep, broken power-law emissivity profiles can be produced in
models with a source positioned close to the symmetry axis (Fig.\
\ref{fig:f1}(a,b); see also Martocchia, Karas \&  Matt 2000), and
indeed these models were invoked to explain the broad line profile  in
MCG--6-30-15 (e.g.\ Miniutti et al.\ 2003).  We note that in these
models the source height  has to be rather small if the steep
emissivity is to extend beyond the ergosphere  (see also Martocchia,
Matt \& Karas 2002).  In particular, we find that in model $C$ the
steep  $\epsilon_{\rm Fe}(r)$ profile, with $q>4$, is produced only in
the ergosphere, $r_{\rm d}<2$, if  $h_{\rm s}>1$
(Fig.~\ref{fig:f4}(b)). However, photons emitted from within $r_{\rm
d}=2$ cannot produce the extended red wing observed in MCG--6-30-15,
since they are redshifted by $1/g=2$--3 (for inclination $\theta_{\rm
obs}=30^{\circ}$), and so they are registered  at $E<3.5$ keV (their
contribution is shown in Fig~\ref{fig:f7}(d)).  Fe lines resulting
from $h_{\rm s}>1$ do have profiles skewed towards low energies,
however they do not posses elongated red wings with approximately
constant slope between 3.5 and 5.5 keV, as observed in
MCG--6-30-15. For example, for $h_{\rm s} = 3-8$, as required by the
variability model of Miniutti et al.\ (2003), the profile is much less
pronounced than  that  corresponding to parameters of the best fit
from Fabian et al.\ (2002) (Fig~\ref{fig:f7}(b,c));  compare with
fig.\ 5 in Miniutti et al.~2003).  We find that to fit the profile
observed in MCG--6-30-15 model $C$ requires the source height $h_{\rm
s} < 1$, see  Fig.\ \ref{fig:f7}(f).

Similar constraints are obtained in model $A$. Here, in order to
produce steep emissivity ($q > 4$) at  $r_{\rm d}>r_{\rm erg}$, the
source must be located at  $r_{\rm s}  \la 4$.  Explaining the red
wing in MCG--6-30-15 requires  $r_{\rm s} \la 2$,  with the best
agreement achieved for $r_{\rm s}=1.6$, Fig.\ \ref{fig:f7}(d). This
overproduces somewhat the flux below 3.5 keV, if $r_{\rm in}<2$ (hence
it requires $r_{\rm in} \simeq 2$, in agreement with Fabian et al.\
2002).

Miniutti et al.\ (2003) claim that their model fits the data for
$h_{\rm s} = 3 - 8$. We note that for $h_{\rm s} > 3$ the flux below 5
keV  is actually weaker than for  the model with a single power-law
with $q=3$, compare the dashed curve in  Fig.\ \ref{fig:f7}(a) and the
solid curve in Fig.\ \ref{fig:f7}(c). Moreover, for $h_{\rm s} = 8$,
the line profile is fully consistent  with the one produced by
emission only from $r_{\rm d}>6$ in a model with $a=0$, see Fig.\
\ref{fig:f7}(b).  Then, the fits in  Miniutti et al.\ (2003) appear to
contradict results of, e.g., Fabian et al.\ (2002) who find that
strong emission from $r_{\rm d}<6$ is required. We suspect that this
discrepancy results from modelling of the continuum emission.   The
simultaneous observation with {\it Beppo-SAX}, taken into account by
Fabian et al.\ (2002), allows to constrain the underlying power-law
continuum and the Compton reflected component. On the other hand,
Miniutti et al.\ (2003)  use only data in 3--10 keV range, which
poorly constrains the continuum. Moreover, the fitted value of
$\Gamma$ is not given in their paper so it is not  clear if it is
consistent with data in broader energy range. Presumably, softer
power-law emission may compensate the red wing. We stress, however,
that if the model of Miniutti et al.\ (2003) with, e.g.,  $h_{\rm s} =
8$, could fit the data in the broader energy range, it would have an
essential implication, namely, rotation of the black hole in
MCG--6-30-15 would not be required by the time-averaged profile from
the {\it XMM-Newton} observation.

\begin{figure}
\centerline{\includegraphics[height=43mm]{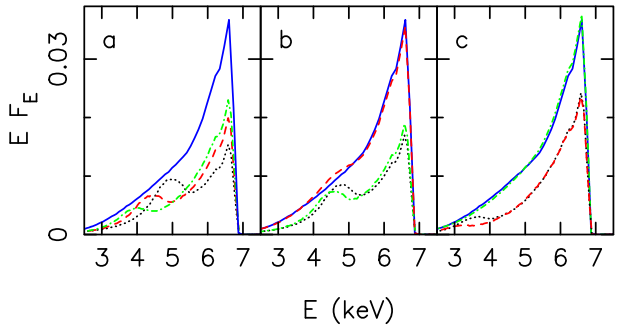}}
\caption{The figure illustrates formation of a profile for
superposition  of primary X-ray sources in model $S_{\rm K}^{\rm PT}$
with $I(\mu_{\rm em}) \propto 1 + 2.06 \mu_{\rm em}$. The profiles for
superpositions are rescaled for clarity. The solid (blue) curve in all
panels shows the superposition profile  for $r_{\rm s}=1.8$, 2, 2.2,
2.5 and 3 ($D_1$).  {\bf (a)}  The dotted (black), red (dashed) and
dot-dashed (green) curve  show individual profiles for $r_{\rm s}=3$,
2.5 and 2.2. {\bf (b)} The dashed (red) curve shows the superposition
profile for  $r_{\rm s}=1.8$, 2, 2.2, 2.4, 2.6, 2.8 and 3 ($D_2$). The
dot-dashed (green) and dotted (black) curve show individual profiles
for $r_{\rm s}=2.6$ and 2.8. {\bf (c)} The dot-dashed (green) curve
shows the superposition profile  for sources  with $r_{\rm s}=1.5$,
1.7, 1.8, 2, 2.2, 2.5 and 3. The dashed (red) and dotted (black)
profiles are for $r_{\rm s}=1.8$ and 2, respectively.}
\label{fig:f35}
\end{figure}

In general, enhanced illumination of  the disc between $r_{\rm d}=2$
and 6, required to model the strong, elongated red wing, seems to
require the primary emission concentrated close to this area of the
disc, as in models $A$ and $C$ with small $r_{\rm s}$ or $h_{\rm s}$.
Similarly, such an enhanced illumination occurs in model $S_{\rm K}$
with small $r_{\rm s}$.  For a single primary source, in model $S_{\rm
K}$, the best agreement with parameters from Fabian et al.\ (2002) is
achieved for $r_{\rm s} \la 2$, Fig.\ \ref{fig:f7}(e). For larger
$r_{\rm s}$, the model predicts profiles possessing the red tail with
a pronounced bumpy feature (cf.\ Figs.\ \ref{fig:f6}(b,c)), which is
not seen in the time-averaged profile of the observed line.   On the
other hand, superposition of profiles corresponding to various $r_{\rm
s}$ smooths out the shape of the red wing, see
Fig.~\ref{fig:f35}. Then, we explore below in some details a
model with variability effects resulting primarily from the change of
the distance, $r_{\rm s}$, at which dominant contribution of hard
X-rays is produced and with the time-averaged spectrum formed by
superposition of spectra corresponding to various  $r_{\rm s}$.

Motivated by variability effects, discussed in the next section, we
focus on model $S_{\rm K}^{\rm PT}$ and we constrain the range of
distances to $r_{\rm s} \le 3$, which  range is relevant to modelling
the variability. Moreover, significant contribution from $r_{\rm s}
\ga 3$ would yield the blue peak broader than the observed  one, at
least for limb-darkened $I(\mu_{\rm em})$, see below. The detailed
shape of the superposition profile depends (weakly for the considered
range  of $r_{\rm s}$, see below) on relative magnitudes of
contribution from various  $r_{\rm s}$. These, in turn,  result from
both the dependence of  intrinsic luminosity on $r_{\rm s}$, which is
fixed in model $S_{\rm K}^{\rm PT}$,  and the length of time at which
emission is generated at various $r_{\rm s}$.  The latter should
correspond to the length of periods spent by the source  in various
flux states.

The light curve for the 2001 observation (figure 12 from Vaughan \&
Fabian 2004)  indicates that the source spends most of the time in
intermediate flux states, with rare transitions to the very low state
and more frequent, but very short, periods of higher flux states.   We
identify the lowest flux states with $r_{\rm s}<1.8$ (see Section
\ref{sec:low}) and the highest flux states with $r_{\rm s} \approx 3$.
Then, the most frequent, intermediate flux states would correspond to
$r_{\rm s} \simeq 2-2.4$.  These would  yield the difference by a
factor of $\sim 1.5$  between direct fluxes of primary emission  in
the intermediate and high flux states.

\begin{figure}
\centerline{\includegraphics[height=70mm]{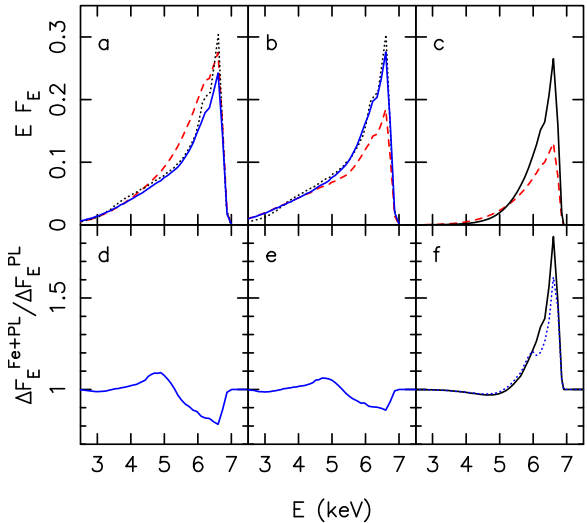}}
\caption{The dotted (black) curves in (a) and (b) show profile of the
Fe line with  parameters of the best fit from Fabian et al.\ (2002),
same as in Fig.\ \ref{fig:f7}.  {\bf (a)} The solid (blue) and dashed
(red) curve shows the superposition profile for primary X-ray sources
distributed in the range $r_{\rm s}=1.8 - 3$  ($D_1$; same as the
solid curves in Fig.\ \ref{fig:f35}) and $r_{\rm s}=1.8 - 20$,
respectively,  in model $S_{\rm K}^{\rm PT}$ with $I(\mu_{\rm em})
\propto 1 + 2.06 \mu_{\rm em}$.  {\bf (b)} The solid (blue) and dashed
(red) curve shows superposition profile for primary X-ray sources
distributed in the range $r_{\rm s}=1.8 - 3$ ($D_1$) in model $S_{\rm
K}^{\rm PT}$ with $I(\mu_{\rm em}) \propto \log (1 + 1/ \mu_{\rm
em})$.  The solid curve includes additional contribution from the
source located close to the axis  at $r_{\rm s}=10$ (model [$S_{\rm
K}^{\rm PT}$,$A$], see text).  {\bf (c)} The solid and dashed curve
shows profile for $h_{\rm s}=8$  and 4, respectively, in model $C$
with $I(\mu_{\rm em}) \propto 1 + 2.06 \mu_{\rm em}$.  The bottom
panels show the difference spectra corresponding to models shown in
top panels.  {\bf (d)} The difference between spectra for $r_{\rm
s}=3$ and $r_{\rm s}=1.8$ in model $S_{\rm K}^{\rm PT}$ with
$I(\mu_{\rm em}) \propto 1 + 2.06 \mu_{\rm em}$.  {\bf (e)} Same as in
(d) but for $I(\mu_{\rm em}) \propto \log (1 + 1/ \mu_{\rm em})$.
{\bf (f)} The solid (black) curve shows the difference between spectra
for $h_{\rm s}=8$ and 4 in model $C$ with $I(\mu_{\rm em}) \propto 1 +
2.06 \mu_{\rm em}$.  The dotted (blue) curve is for $I(\mu_{\rm
em})=I_0$ and $r_{\rm out}=100$ as in Miniutti et al.\ (2003).}
\label{fig:f17}
\end{figure}

As illustrated in Fig.~\ref{fig:f35}, the detailed shape of the line
is weakly  sensitive to specific distribution of $r_{\rm s}$, provided
that the distribution involves a few positions within $r_{\rm
s}=2.2-3$ (contributions from these larger $r_{\rm s}$ are shown in
panels (a) and (b)), so that the red wing is smoothed out. In next
sections  we consider model $S_{\rm K}^{\rm PT}$ involving $r_{\rm
s}=1.8$, 2, 2.2, 2.5 and 3 (denoted below as $D_1$), for which the
superposition profile reproduces the observed line, see
Fig.~\ref{fig:f17}(a). Furthermore, this distribution of $r_{\rm s}$
approximates the distribution of flux states indicated by the
light-curve, with  higher fluxes less frequently sampled, more closely
than a uniform distribution of $r_{\rm s}$ (denoted as $D_2$). Note,
however, that for the uniform distribution $D_2$ only a weak excess
between 4 and 5 keV occurs with respect to $D_1$, see
Fig.~\ref{fig:f35}(b).  These two distributions of $r_{\rm s}$ yield
also similar equivalent widths (EW$_{\rm tot}$, see Section
\ref{sec:ew}) of the Fe line, namely 670 and 620 eV for $D_1$ and
$D_2$, respectively.

Individual profiles of lines for small $r_{\rm s}$ ($\le 2$)
differ weakly from each other as well as from the profile of the
superposition, see  Fig.~\ref{fig:f35}(c).  Therefore, the
superposition profile is insensitive to details of contribution from
small $r_{\rm s}$, e.g., the profiles for $r_{\rm s}=1.5-3$ and
$1.8-3$ are almost identical (Fig.\ \ref{fig:f35}(c)). This property
is more thoroughly explored in Section \ref{sec:low}.

Superposition of lines from sources located at $r_{\rm s} \le 3$
reproduces shape of the observed line at $E < 5$ keV (see
Fig.~\ref{fig:f17}(a,b)) for both limb-darkened and limb-brightened
emission.   On the other hand, details of modelling the line at higher
energies depend on angular emissivity law.  The strong blue peak, at
$E > 5.5$ keV,  has to be formed by photons originating mostly  from
the inner disc rather than from a distant material.  Vaughan \& Fabian
(2004) find that the width of the peak requires the emitting region to
be located at $r_{\rm d} \sim 30$. However, they use a
non-relativistic formula for the Doppler effect, therefore that value
gives an upper limit on the distance of that  region. With a fully
relativistic modelling,  we find that the peak may be reproduced with
the major contribution coming from $r_{\rm d} \sim 10$.  For
$I(\mu_{\rm em}) \propto 1 + 2.06 \mu_{\rm em}$, illumination of that
region by sources located within  $r_{\rm s} =1.8-3$ is approximately
sufficient to yield  the peak with the observed strength, the solid
curve in Fig.\ \ref{fig:f17}(a).  On the other hand, for this
$I(\mu_{\rm em})$, strong emission from $r_{\rm s} =4-10$  would
overproduce the line flux in 5-6 keV range, yielding too large width
of the line core, the dashed curve in Fig.\ \ref{fig:f17}(a).

For $I(\mu_{\rm em}) \propto \log (1 + 1/ \mu_{\rm em})$,
primary sources described by model $S_{\rm K}^{\rm PT}$ are not
sufficient to reproduce the strength of the blue peak, see the dashed
curve  in Fig.\ \ref{fig:f17}(b). In this case, an additional primary
source illuminating the disc mostly at $r_{\rm d} \simeq 10$, e.g. a
source at $r_{\rm s} \simeq 10$ at the symmetry axis, is required  to
reproduce the blue peak. In next sections we discuss some properties
for the specific case of sources described  by model $S_{\rm K}^{\rm
PT}$ with $r_{\rm s} =1.8-3$ and an additional source, described by
model $A$ with $r_{\rm s} = 10$ and $V=0$,  emitting 10 per cent of
total X-ray luminosity. This model, denoted below as [$S_{\rm K}^{\rm
PT}$,$A$], yields the time-averaged profile in very good agreement
with the observed one, see the solid curve in Fig.\ \ref{fig:f17}(b).

\subsection{Variability}

A model consisting of a strongly variable, in normalisation, power-law
and  a more constant reflection component explains variability of
MCG--6-30-15  measured by  {\it ASCA} (Shih et al.\ 2002) and {\it
XMM-Newton} (e.g., Fabian \& Vaughan 2003). Specifically, figure 5 in
Fabian \& Vaughan (2003) shows changes of magnitude of reflection by a
factor of $\la 2$ corresponding to changes of the primary radiation by
a factor of 5.

Such reduction of variability  could be entirely explained by GR
effects for a source, with constant intrinsic luminosity, described by
model $S_{\rm K}$ and changing the radial distance within $r_{\rm s} =
2-3$,  see the solid curve in Fig.\  \ref{fig:f8}(a). In this case,
however, variations of the line, although reduced, would be correlated
with those  of the primary radiation.  On the other hand, Fabian \&
Vaughan (2003) do not reveal any such trend in variations of both
components. Then, it seems that modelling of the observed variability
should involve changes of the intrinsic luminosity with $r_{\rm
s}$. Specifically, model $S_{\rm K}^{\rm PT}$ with $r_{\rm s} =
1.8-3$, yielding approximately constant reprocessed flux for the
primary flux varying by a factor of 3, appears to be the most feasible
to explain that  variability pattern.

Additional changes, by a factor of $\la 2$,   of the intrinsic
luminosity at a given $r_{\rm s}$ would be  required in model $S_{\rm
K}^{\rm PT}$ to account for the spread of reprocessed flux values for
a specific value of the primary flux.   Alternatively, such variations
of the reprocessed component, at fixed primary component, could be
explained by changes of  $h_{\rm s}$, as for $h_{\rm s}<1$ such
variations of the height yield an almost constant primary flux and the
Fe flux varying by a factor of 2, see the dotted curve in Fig.\
\ref{fig:f8}(a).

The reduction effect is weakened if even a small fraction of X-rays is
generated at larger distances, as required by the time-averaged
profile for limb-brightened $I(\mu)$, see Fig.\
\ref{fig:f17}(b). E.g.,  for model [$S_{\rm K}^{\rm PT}$,$A$],
involving  a constant source, located at $r_{\rm s}=10$ at the axis
and emitting    10 per cent of total X-ray luminosity, variations of
the distance of a source described by $S_{\rm K}^{\rm PT}$ within
$r_{\rm s}=1.8-3$ yield the change the observed primary flux by a
factor of 2 (at the constant Fe flux).

A similar reduction, by a factor of 2, of the Fe flux variations is
predicted in  model $A$   as well as in model $C$,  interestingly, in
both cases in the range of parameters  relevant for modelling the
time-averaged profile.  Namely, in model $A$, this effect - involving
enhancement by returning radiation, see Section \ref{sec:var:a} -
occurs for small $r_{\rm s}$, for which the line with strong red wing
is produced, cf.~Fig.\ \ref{fig:f7}(d).  In model $C$, reduced
variations of the Fe flux - basing on specific $V(h_{\rm s})$ - occur
for the intermediate heights, for which the model fits the observed
profile according to Miniutti et al.\ (2003).  On the other hand,
neither of these models is able to explain  some finer details of
spectral variability revealed in the  rms spectra, see below.  Note
that effects related to light-travel time are (at most) marginally
important in these applications of models $A$ and $C$ to MCG--6-30-15.
In particular, for $h_{\rm s}<8$ in model $C$, fraction of Fe photons
from $r_{\rm d}>150$ is smaller than 5 per cent and  the longest delay
time (corresponding  to a small fraction of photons coming from the
farthest, to observer, side of the disc) for photons reflected from
$r_{\rm d} = 150$ is smaller than 3.5 ks (for $M = 3 \times 10^6
M_{\sun}$).

Note also that the above is not the unique variability pattern
observed in MCG--6-30-15. Namely, Reynolds et al.\ (2004) find that
during the  2000 {\it XMM} observation, for the range of the 2-10 keV
flux between $F_{2-10} = 2 \times 10^{-11}$ (the deep minimum level)
and  $4 \times 10^{-11}$ erg/s/cm$^2$ (typical state), the line
followed variations of the power-law component, keeping approximately
constant equivalent width. This is clearly a different behaviour than
during the {\it ASCA} as well as the 2001 {\it XMM-Newton}
observations when, in the same range of $F_{2-10}$, the line did not
vary, leading to strongly reflection dominated spectra at low-flux
states (e.g. Iwasawa 1996, Fabian \& Vaughan 2003, see also figs.\ 13
and 14 in Vaughan \& Fabian 2004).  This indicates that the nature of
the physical mechanism underlying variability during the 2000
observation  is different than that of other observations. A plausible
interpretation involves assumption that, in the former, variability
was primarily due to variations of intrinsic luminosity while in other
cases it was dominated by GR effects.

The rms spectra for the 2001 {\it XMM} observation reveal the
suppressed variability around 6.5 keV,  see e.g.\ figure 16 in Vaughan
\& Fabian (2004).  Such suppression could be due to the presence of
reflection from distant material, as suggested for explanation of
similar features in  the rms spectra of several other objects (see,
e.g., Ponti et al.\ 2006).  In MCG--6-30-15, however, contribution of
such a constant spectral component is very weak (see Lee et al.\ 2002,
Iwasawa et al.\ 1996).  Therefore, the suppressed  variability of the
blue peak of the line should be explained within the GR variability
model.  As discussed in Section \ref{sec:rms}, only model $S_{\rm
K}^{\rm PT}$ (or $S_{\rm K}$) is able to explain such property.
Moreover, the decline around 6.5 keV in the rms  spectrum occurs in
the range of $r_{\rm s}$ for which the model reproduces  the
time-averaged profile.
  
On the other hand, Vaughan \& Fabian (2004) indicate hints for
increased variability around the Fe line energy in the rms spectrum
for lower flux states, although ambiguities remain in interpretation
of that analysis (see their  figures 16 and 17 and discussion in
section 5.3.2).  We emphasise that such property, if confirmed, would
contradict predictions of model $S_{\rm K}^{\rm PT}$, where a slightly
stronger decline in the rms spectrum should correspond to lower
luminosity states, see Fig.\ \ref{fig:f19}. Then, in the framework of
model $S_{\rm K}^{\rm PT}$, such property would require stronger
variations of intrinsic luminosity,  at a fixed $r_{\rm s}$, for
smaller $r_{\rm s}$.

Ponti et al.\ (2004) note an excess of fractional variability in the
4.7-5.8 keV band (followed by a drop at energies close to 6.4 keV) in
the rms spectrum of the 2000 observation,  suggesting presence of a
spectral component   which varies more than the underlying continuum.
Such an excess may result from variations of the red wing of the line
corresponding to changes of the position within $r_{\rm s}=2-4$,   see
Fig.\ \ref{fig:f15}(c).

\subsection{Time-resolved profiles; difference spectra}

High-quality spectra for time-scales on which the flux remains
approximately constant should reveal line profiles corresponding to a
fixed position of the X-ray source. Specifically, for model $S_{\rm
K}^{\rm PT}$ these profiles should exhibit clear bumpy features in the
red wing.  MCG--6-30-15 varies strongly on time-scales $<1$ ks.
Current data allow to determine spectra for time bins of at least 10
ks. Then, these spectra would involve averaging over certain ranges of
$r_{\rm s}$ in model $S_{\rm K}^{\rm PT}$, which in turn would  smooth
out the red wing (see also Fig.\  \ref{fig:f21}(c)).   We note,
however, clear indications of bumps in the line profile below $6$ keV
in the {\it ASCA} observations and some, less significant, hints in
the {\it XMM-Newton} data. If interpreted in terms of  model $S_{\rm
K}^{\rm PT}$ predictions,  this would indicate that in some cases
generation of X-rays  is localised around specific $r_{\rm s}$ on
timescales $\ge 10$ ks.

The time-averaged profile of 1999 {\it ASCA} observation (see figure 1
in Shih et al.\ 2002)  has a main peak at $\sim$6.5 keV and  then an
additional peak at $\ga$5 keV (qualitatively similar but less
pronounced peaks around 5 keV are seen in time averaged profiles from
1994 and 1997 {\it ASCA} observations; see Tanaka et al.\ 1995 and
Iwasawa et al.\ 1999).    This profile is modelled with a rather flat
radial emissivity, $q \la 3$, and a small outer radius, in the range
$r_{\rm out}=10$-30.  For such parameters,  the lower energy peak is
formed in the way illustrated  in Fig.\ 11(c).  However, that fitting
model requires some complex geometry of the central region, with
obscuration of more distant parts of the disc, justifying the above
values of   $r_{\rm out}$; otherwise,  even  a single point-like
source in the central region would illuminate more distant regions of
the disc,  giving rise to a rather smooth profile, without the peak at
5 keV.  On the other hand, the observed profiles may be explained by
increased X-ray activity at $r_{\rm s} \approx 3$ in model $S_{\rm
K}^{\rm PT}$, without assuming such small  values of $r_{\rm out}$.

Such clear features are not revealed by {\it XMM-Newton} for similar
flux states. However, the time-resolved (in 10 ks bins) spectra often
show some   features at (2--3)$\sigma$ level around 3.5--4 keV in the
2000 observation (see figure 8  in Reynolds et al.\ 2004; note that
the time-average profile in that figure has a clear bump  with maximum
at $\la 4$ keV) and around 4.5--5 keV in the 2001 observation (see
figure 18 in Vaughan \& Fabian 2004).  Finally, Dov\v{c}iak et al.\
(2004a) note features near  $E \sim 4.8$ keV in the time-averaged line
profile of 2001 observation,  which could be interpreted as a Doppler
horn resulting from locally enhanced emission on some part of the
disc. The above energies would correspond to $r_{\rm s} \approx 2$ and
3 for the 2000 and 2001 observation, respectively.

Given a rather poor quality of the time-resolved spectra, less direct
approaches  to investigate spectral changes are made.  Specifically,
the difference spectra between various flux levels are determined,
which show only the variable component of the spectrum, while the
non-variable component is subtracted away. Figure 15 in Vaughan \&
Fabian (2004) shows the difference spectrum between the average
spectra, for the whole 2001 observation, with fluxes  lower and higher
than the mean.  The difference is consistent with a power-law, with
residuals from the power-law model not exceeding 10 per cent.
Therefore, it indicates very weak change of the  spectrum of the
reprocessed component. This property strongly challenges the GR models
of variability, which in general predict rather significant changes of
the line profile resulting from the  change  of the source
position. Below we present  example difference spectra for models
$S_{\rm K}^{\rm PT}$ and $C$.

We note, however,  that both the low and high flux spectra in the
analysis in Vaughan \& Fabian (2004) involve averaging over   wide
ranges of flux states, see their figure 12, and the average fluxes for
high and low flux spectra differ by a factor of 1.5.  In GR
variability models, such averaging would correspond to averaging over
large ranges of $r_{\rm s}$.  Taking into account that, e.g.,  model
$S_{\rm K}^{\rm PT}$ should involve variations of intrinsic luminosity
by a factor of $\la 2$, these ranges of $r_{\rm s}$ may largely
overlap.  The most interesting constraint on spectrum presumably
corresponding to an  approximately fixed position is discussed in
Section \ref{sec:low}.

Figs.\ \ref{fig:f17}(d-f) show difference between spectra   for
$r_{\rm s}=3$ and $r_{\rm s}=1.8$ (i.e.\ between extreme positions
considered above for the time-averaged profile) in model $S_{\rm
K}^{\rm PT}$ and for $h_{\rm s}=8$ and $h_{\rm s}=4$ in model $C$.
The difference between the total spectra (Fe line + power-law) is
divided by the difference  of the primary power-laws.  Variable red
wing gives rise  to an excess at $\la 5$ keV, not exceeding 10 per
cent. The exact energy of this excess depends on the value of $r_{\rm
s}$ attributed to the high state, with higher energies corresponding
to larger  $r_{\rm s}$.  Owing to effects discussed in Section
\ref{sec:var:s}, model $S_{\rm K}^{\rm PT}$ is characterised by
relatively small variations of the blue peak. Deviations of the
difference spectrum from a power-law, at the peak energy, depend on
the angular emissivity law, with limb-brightened laws giving rise to
smaller deviation.

In models $A$ and $C$ significantly stronger variations of the blue
peak occur,  therefore, these models predict much larger residuals
around 6 keV (than model $S_{\rm K}^{\rm PT}$) in the difference of
spectra between positions corresponding to low and high fluxes, see
Fig.\ \ref{fig:f17}(f).   Note also that significant deviations from
power-law may occur in the difference    between two profiles which
fit the same observation equally well, see the dotted curve in Fig.\
\ref{fig:f17}(f), cf.\ Miniutti et al.\ (2003).

The difference spectrum between the 2001 and 2000 {\it XMM}
observations shows an excess at 6.6 keV, see figure 20 in Vaughan \&
Fabian 2004,  which could be due to a change in the flux of  the blue
peak of the Fe line. Such a change of the blue peak, yielding appropriate 
level of residuals around 6.6 keV, could result from significant change of
the source height, as illustrated in Fig.\ \ref{fig:f17}(f), with
the 2001 observation corresponding to larger $h_{\rm s}$. On the other
hand, this interpretation would not be consistent with other
properties of the 2001 observation, discussed in this paper, which
seem to require small $h_{\rm s}$. Moreover, the line in the
difference  between the 2001 and 2000  observations is narrow, while
that resulting from the change of $h_{\rm s}$ would be rather broad,
with significant flux below 6 keV. 

\subsection{Equivalent width}
\label{sec:ew}

High values of both the equivalent width (EW) of the Fe line   and the
normalisation parameter of Compton reflection  ($R=\Omega/2 \pi$),
characterising the X-ray spectrum of MCG--6-30-15, are supposed to
result from enhancement of disc irradiation by GR effects.  Slightly
different values of EW for the 2001 observation result from the two
{\it XMM-Newton} detectors.  Specifically, using the same model  (with
$R=1$), Fabian et al. (2002) and Fabian \& Vaughan (2003) find EW =
550 eV and EW = 685 eV using the data from MOS and pn camera,
respectively.  Allowing for higher normalisation of reflection,
$R=2.2$,  Fabian et al. (2002)  obtain EW =  410 eV.  Using the {\it
BeppoSAX} data, Ballantyne, Fabian \& Vaughan (2003) find a high value
of reflection parameter, $R>2.6$, in agreement with the above EW
values.

Note that procedure of computing the EW is particularly important  in
GR models. Specifically, for high $R$, the EW with respect to the sum
of the power-law and Compton reflection, denoted below as EW$_{\rm
tot}$,  may be much smaller than the EW with respect  to the power-law
only, denoted as EW$_{\rm PL}$ (see also examples of EW values for
various procedures of computing in Miniutti et al.\ 2004).

\begin{figure}
\centerline{\includegraphics[height=47mm]{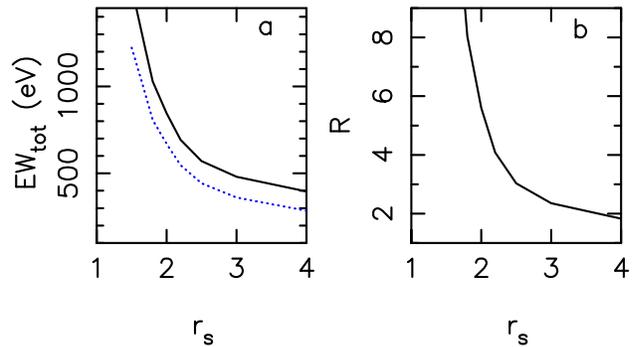}}
\caption{The solid curve in panel (a) and (b) shows the line EW (with
respect  to power-law + reflection continuum) and the reflection
parameter, respectively, as a function of the source distance $r_{\rm
s}$, for  $S_{\rm K}^{\rm PT}$ with $I(\mu_{\rm em}) \propto 1 + 2.06
\mu_{\rm em}$.  The reflection parameter is determined as $R=$EW$_{\rm
PL}$/EW$_{\rm PL}^{\rm class}$, see text, with EW$_{\rm PL}^{\rm
class}=260$ eV.  The dotted (blue) curve is for $I(\mu_{\rm em})
\propto \log (1 + 1/ \mu_{\rm em})$}
\label{fig:f33}
\end{figure}

We have determined both EW$_{\rm tot}$ and EW$_{\rm PL}$ (all values
given below  correspond to $\theta_{\rm obs} = 30\degr$). The latter
may be used, by comparing with the classical value  (i.e.\ for the
model with no relativistic effects), EW$_{\rm PL}^{\rm class}$,  to
estimate the reflection normalisation parameter as  $R=$EW$_{\rm
PL}$/EW$_{\rm PL}^{\rm class}$.  The  value of EW$_{\rm PL}^{\rm
class}$ depends on the angular emission law, in accordance with
changes of the line profile discussed in Section \ref{sec:limb}.
E.g., EW$_{\rm PL}^{\rm class}=170$ and 260 eV corresponds to
$I(\mu_{\rm em}) \propto \log (1 + 1/ \mu_{\rm em})$ and $I(\mu_{\rm
em}) \propto 1 + 2.06 \mu_{\rm em}$, respectively. Neglecting the
factor 1.3 (related with elemental abundance, see Section 2), we find
EW$_{\rm PL}^{\rm class}$=130 eV (in agreement with George \& Fabian
1991; see their figure  14)  for $I(\mu_{\rm em}) \propto \log (1 + 1/
\mu_{\rm em})$. However, dependence of  $R$ on specific $I(\mu_{\rm
em})$ is rather weak; differences between limb-brightened and
limb-darkened cases do not exceed 20 per cent.  In Fig.\
\ref{fig:f33}(b) we show values of $R$  corresponding to limb-darkened
$I(\mu_{\rm em})$, which is closer  to angular dependence of Compton
reflected component.

In model $S_{\rm K}^{\rm PT}$ with small $r_{\rm s}$, the  EW$_{\rm
tot}$  is extremely sensitive to $r_{\rm s}$, see Fig.\
\ref{fig:f33}(a). For $I(\mu_{\rm em}) \propto 1 + 2.06 \mu_{\rm em}$,
the EW$_{\rm
 tot}$ increases from  480 to 1030 eV with the distance
decreasing from $r_{\rm s}=3$ to 1.8. The superposition of sources
within $r_{\rm s}=1.8-3$ ($D_1$, the solid curve in Fig.\
\ref{fig:f17}(a)) yields EW$_{\rm tot}$ = 670 eV. For $I(\mu_{\rm em})
\propto \log (1 + 1/ \mu_{\rm em})$, in the same range of  $r_{\rm
s}$, EW$_{\rm tot}$ increases from 360 to 800 eV, and the
superposition  yields EW$_{\rm tot}$ = 520 eV. Then, for the
limb-brightened $I(\mu_{\rm em})$, EW$_{\rm tot}$  is in approximate
agreement  with spectral fits of the {\it XMM-Newton} data. The
limb-darkened $I(\mu_{\rm em})$  yields a slightly higher value,
although this conclusion is affected by uncertain abundance of iron.

The reflection parameter for the superposition of sources within
$r_{\rm s}=1.8-3$ in model $S_{\rm K}^{\rm PT}$ is $R = 3.8$, which
value  is within the confidence  limit from analysis of the {\it
BeppoSAX} data in Ballantyne et al.\ (2003).

The EW is  significantly reduced if even a small fraction  of X-rays
is emitted at larger distances, as required by the time-averaged
profile for limb-brightened  $I(\mu_{\rm em})$. E.g, an additional
source  at the axis, $r_{\rm s}=10$, in model [$S_{\rm K}^{\rm
PT}$,$A$] (the solid curve in Fig.\ \ref{fig:f17}(b)), emitting 10 per
cent of the  X-ray luminosity, yields  EW$_{\rm tot}$ = 380 eV and $R
\la 3$.

\subsection{X-ray luminosity}

The primary emission, observed at small  $\theta_{\rm obs}$, is
strongly reduced in model $S_{\rm K}^{\rm PT}$ for the range of
distances relevant to modelling of the line profile and variability
effects. In particular, for $r_{\rm s}=3$ and 1.8, it is reduced by a
factor of 8 and 70, respectively. Then, the observed hard X-ray flux
would be reduced by over an order of magnitude due to GR effects.  The
2--10 keV flux in the typical state of MCG--6-30-15 corresponds to
isotropic luminosity $L_{2-10}^{\rm iso}=4 \times 10^{42}$ erg/s,
implying the total isotropic luminosity  $L_{\rm X}^{\rm iso} \la 2
\times 10^{43}$ erg/s.  The above reduction factors imply that the
intrinsic luminosity of the X-ray source is $L_{\rm X} \simeq 2 \times
10^{44}$ erg/s $\approx 0.5 L_{\rm Edd}$ (for $M=3 \times 10^6
M_{\sun}$)  and this in turn indicates  $\dot M \ga \dot M_{\rm Edd}$,
making the central engine of  MCG--6-30-15 closely similar to those of
Narrow Line Seyfert 1 galaxies (e.g. Collin \& Kawaguchi 2004).

The NIR/optical/UV luminosity (presumably thermal emission of an
optically thick disc) is estimated to $L_{\rm disc} \ge 2 \times
10^{43}$ erg/s by Reynolds et al.\ (1997). The above lower limit
corresponds to the minimum reddening allowed for MCG--6-30-15.  The
putative rapid rotation of the black hole implies that the thermal
disc emission, driven either by direct dissipation of the accretion
power or by irradiation by X-rays, should emerge mostly from distances
of a few $R_{\rm g}$. Then, the disc emission would be subject to
similar reduction as the X-rays and therefore the disc luminosity may
be comparable to that of the X-ray source (or higher for larger values
of the reddening).  Actually, this scenario, i.e. a more luminous disc
emission beamed along the equatorial plane, is consistent with very
high luminosity of MIR/FIR radiation coming from reprocessing of
central emission, see discussion in Reynolds et al.\ (1997).

\begin{figure}
\centerline{\includegraphics[height=70mm]{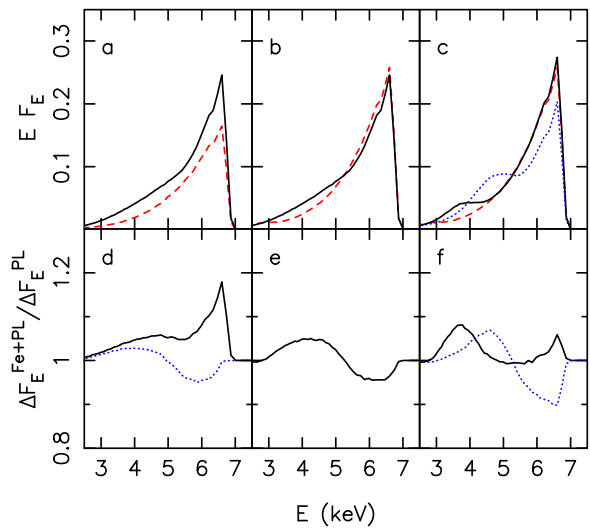}}
\caption{The figure shows the difference between spectra of higher
flux states, averaged over some ranges of source position, and the
lowest flux state, for a fixed source position,  in model $S_{\rm
K}^{\rm PT}$ with $I(\mu_{\rm em}) \propto 1 + 2.06 \mu_{\rm em}$.
The solid (black) curves in panel (a) and (b) show superposition of
profiles for  $r_{\rm s}=1.8 - 3$, same as the solid curve in Fig.\
\ref{fig:f17}(a). The solid (black) and dotted (blue) curve in (c)
shows superposition for  $r_{\rm s}=2-2.2$ and $2.5-3$, respectively.
The dashed (red) curves show profiles for $r_{\rm s}=1.5$ (a) and
$r_{\rm s}=1.7$ (b,c).  Normalisations of spectra for the
superposition and for the single source position  are adjusted so that
they correspond to equal times of observation.  The solid (black)
curves in panels (d-f) and the dotted (blue) curve in (f) show the
difference spectra (the lower flux spectrum subtracted from the higher
flux spectrum) for parameters indicated for the top panels.  The
dotted (blue) curve in panel (d) is for the same parameters as the
solid one but with the spectrum for $r_{\rm s}=1.5$ multiplied by a
factor of 1.5.}
\label{fig:f21}
\end{figure}

\subsection{Low flux state}
\label{sec:low}

The lowest flux states, with spectra dominated by the reflection
component, are particularly interesting for studying the strongly
relativistic regime of space-time.  In the GR models, these states
would correspond  to positions of the X-ray source in the immediate
vicinity  of the event horizon.  Properties of the lowest flux spectra
seem to deviate between various observations of MCG--6-30-15.

The line profile of the deep minimum state analysed by Iwasawa et al.\
(1996) is completely dominated by the red wing, while the blue peak is
strongly suppressed. This could be interpreted as the result of
generation of primary emission at small $r_{\rm s}$ close to the axis
(i.e.\ in the regime of model $A$ rather than $S$), however, some
challenge for this model  results from enhancement  of the blue peak
by illumination of the surrounding disc (i.e. beyond a few $R_{\rm
  g}$) by returning radiation.

Indeed, such a line profile, dominated by the red wing, can be
  modelled with very steep $\epsilon_{\rm Fe}$ profiles ($q>4$) on the
  whole disc, so that contribution from beyond a few $R_{\rm g}$ is
  negligible. As a result, the line is formed in bulk by strongly
  redshifted photons and has a bumpy shape without a blue peak. A
  similar (more extreme) effect is considered as the reason of
  non-detection of a broad line in some spectra presumed to be
  affected by strong GR effects, e.g. NGC 4051 (see e.g.\ figure 7 and
  the related discussion    in Miniutti \& Fabian 2004).    In
  principle, such lines may result from {\it direct} illumination by a
  source, with $V=0$, at a very small $r_{\rm s}$ close to the
  symmetry axis.  However, in any such model, additional illumination
  by the returning radiation would result in generation of a rather
  pronounced blue peak.  The above effect does not occur if the
  reflecting (optically thick) material is absent within at least
  $2R_{\rm g}$.

Reynolds et al.\  (2004) argue that the lack of the blue peak is the
characteristic property of the low state. However, some low flux
profiles do have  strong blue peak, see below, on the other hand,  a
profile  with no blue peak was revealed during the flare in the 1997
{\it ASCA} observation (Iwasawa et al.\ 1999).  Note also that the
analysis in Reynolds et al.\  (2004) involves a wide range of flux
states including much higher than that of the deep minimum. Moreover,
their fits include a narrow Fe line with EW = 100 eV, which value is
much higher than these typically  found for this object (e.g.\ Iwasawa
et al.\ 1996, Lee et al.\ 2002).

Finally, we argue that properties of  the low flux state of the 2001
observation, analysed by  Fabian \& Vaughan (2003), are consistent
with predictions of model $S_{\rm K}^{\rm PT}$. The spectrum of that
state is strongly dominated by reflection, with very large EW = 1.4
keV, similar as for the deep minimum of the {\it ASCA} observation. On
the other hand, the profile has  a strong blue peak, as implied both
by consistency with higher flux profiles, see below, and by the fit
with $q=3.3$ and high $r_{\rm out}$.  To account for such high values
of EW, model $S_{\rm K}^{\rm PT}$ requires  the X-rays to be emitted
from $r_{\rm s} \le 1.7$. For these $r_{\rm s}$, the observed primary
X-ray flux is reduced by a factor of $\ga 100$, in approximate
agreement with normalisation of the power-law component for low flux
state in figure 5 in Fabian \& Vaughan (2003).  Note that for these
$r_{\rm s}$ the strongly irradiated part of the disc, under the
source, does not contribute to the line profile - due to extreme
redshift and reduction of this emission.  Then, the line can be
modelled by a single  radial power-law, with $q \approx 3$, in
agreement with the fit from  Fabian \& Vaughan (2003). Such line
profile does not possess any bumpy features  in the red wing and is
similar to the profile averaged over the range of $r_{\rm s}$
corresponding to higher flux states. This property is consistent  with
those derived from analyses of the 2001 observation, as discussed
below.

The light curve for the 2001 observation in general exhibits rapid
variability, except for a rather prolonged period ($\ga 10$ ks) of the
lowest flux state, see bin 303:k in figure 2 from Fabian \& Vaughan
(2003).  Weak changes of the flux suggest that this period may
correspond to a fixed location of the source. Remarkably, spectrum
from this period is similar to spectrum averaged over some some range
of  higher fluxes; see figure 3 in Fabian \& Vaughan (2003) with up to
20 per cent residuals in the difference spectrum. Similar property is
found in Fabian et al.\ (2002) for  another low flux period; Vaughan
\& Fabian (2004) note that they do not find evidence for strong
systematic residuals  in the difference spectra between the lowest
flux  state and several higher flux levels.

Fig.\ \ref{fig:f21} shows the difference  spectra between the flux
states, averaged over some ranges of $r_{\rm s}$, and the lowest flux
state, predicted in model $S_{\rm K}^{\rm PT}$ for limb-darkened
angular emissivity - which gives rise to the largest deviations in the
difference spectra. For $r_{\rm s}=1.7$, the residuals are small,
approximately within  the  observational constraints.  For very small
$r_{\rm s}$, model $S_{\rm K}^{\rm PT}$ predicts decrease of the line
flux, cf.\ Fig.\ \ref{fig:f8}(c), therefore,  $r_{\rm s}=1.5$ the
residuals are relatively large.  However, the shape  of the line is
similar to the averaged profile and a slight modification  of the
model, yielding higher line flux, reduces the residuals to a small
level, see the dotted curve in Fig.\ \ref{fig:f21}(d).

Note that in two Narrow Line Seyfert 1 galaxies, NGC 4051 and 1H
0707--495, properties of the lowest flux state are different than
predicted by model $S_{\rm K}^{\rm PT}$. In these objects, the
reflected component is correlated with the primary power-law at low
flux values and becomes more constant at higher fluxes  (Fabian et
al.\ 2004, Ponti et al.\ 2006). Similarly as for the 2000 observation
of MCG--6-30-15, significant changes of intrinsic luminosity would be
required to explain those low flux states.

\subsection{Discussion}

Model $S_{\rm K}^{\rm PT}$, with $a=0.998$, seems to be the most
feasible to explain these properties of the X-ray spectrum of
MCG--6-30-15 which are supposed to result from GR effects. Below  we
summarise predictions of the model which are consistent with the
observed  properties
\newline
(i) Primary emission from  sources distributed within $r_{\rm s} =
1.8-3$ gives rise to the line profile possessing a red wing consistent
with that revealed in the 2001 observation.
\newline
(ii) Changes of the distance of the source, within $r_{\rm s} = 1.8 -
3$, yield variations of the  primary emission flux with constant
normalisation of  reflection component.
\newline
(iii) Change of the distance, within $r_{\rm s} = 1.8 - 3$, gives rise
to  a decline in the rms spectrum at 6.5 keV.
\newline
(iv) Primary emission from a source located at $r_{\rm s} \le 1.7$
produces a strongly reflection dominated spectrum, consistent with
that observed in the lowest flux states; moreover, the line profile
for such location is similar to the average profile for larger $r_{\rm
s}$.
\newline
(v) Rather significant changes of the line profile occur
when the distance of a single source changes at $r_{\rm s} > 2$;
the  resulting, systematic residuals in the difference spectra are, however, 
relatively small ($\sim 20$ per cent level) and the model reproduces 
reasonably well a power-law like difference spectrum.
\newline
Some details of the model, e.g.\ modelling of the blue peak  or the EW
value, depend on the angular emission law.

Note that model $S_{\rm K}$, with some modifications, could explain a
smoothed red wing even without superposition, i.e.\ for a single
source  at $r_{\rm s} =2-3$. Moreover, such model could account for
small  changes of the line profile corresponding to the change of
$r_{\rm s}$.  Namely, photons forming the variable, bumpy part of the
red wing may be depleted due to ionization of the disc surface,
neglected in this paper. Qualitatively, the strongest ionization
should occur under the source, where the photons forming the
redshifted horns originate.  We emphasise that if future observations
confirm that the line profile remains strictly unchanged at any flux
level, applicability of model $S$ would indeed require strong
reduction of fluorescent emission from the region under the source,
involving  ionization of that region.

Another possibility to suppress the bumps in the red wing involves
generation of X-rays at  a slightly larger $h_{\rm s}$, for which a
more extended area under the source is illuminated and the wing does
not posses a pronounced feature,  compare Figs.\ \ref{fig:f7}(e) and
\ref{fig:f7}(f).

Our model $S_{\rm K}^{\rm PT}$ represents most closely a model with
flares occurring randomly at various distances. In a realistic
scenario, the X-rays would be always generated by flares distributed
over a range  of $r_{\rm s}$ and the observed Fe profile would
correspond to  the superposition of profiles for a single $r_{\rm s}$.
Variability effects would then result from changes of the distance  of
the strongest activity. These changes of $r_{\rm s}$ giving the
dominating contribution should result in changes of the line profile
described above.   We note, however, a caveat for the model  with
flares. Namely, the model involving compact  sources corotating with
the disc predicts quasi-periodic modulation of the primary emission
because of Doppler effect as the source circles around the centre (see
\.Zycki \& Nied\'zwiecki 2005).  This signal would appear on the
Keplerian time scale, which is shorter than time scales considered in
this paper.  However, such a signal is not observed in the {\it
XMM-Newton\/} data from MCG--6-30-15 (Vaughan, Fabian \& Nandra 2003).
This may imply a continuous spacial distribution of the hard X-ray
source, e.g.\ an extended corona covering the disc surface or a small
hot torus replacing the disc, in both cases within a few innermost
$R_{\rm g}$ (as needed for the variability  effects).

\section{Discussion and conclusions}

\subsection{Variability models}

We have extended the model, formulated by Miniutti \& Fabian (2004),
relating reduced variability of the reflected emission to changing
magnitude  of relativistic effects as location of the primary X-ray
source changes.  We find that original computations of Miniutti \&
Fabian (2004) - for a model involving a vertically moving source -
overestimate  the reduction effect by assuming a value of the outer
radius of the disc which is too small for the range of the source
heights considered in that model.

On the other hand, we find a significant reduction of the variability
of reflected emission in a model with a rapidly rotating black hole
and a source moving radially, low above  the disc surface. The reduced
variability occurs then for the innermost range  of radial distances,
$\le 4R_{\rm g}$.  We find also that - only in this range of
parameters  - the GR effects give rise to  a significant decline
around 6.5 keV in rms spectra.

Note that generation of strong X-ray emission at these distances ($\le
4R_{\rm g}$) is consistent with the condition of a high value of $a$,
as - for a rapidly rotating black hole - most accretion power is
dissipated within a few innermost $R_{\rm g}$.

\subsection{The black hole spin}

Determination of the value of black hole spin  or, even more
fundamentally, verification of effects predicted  by the Kerr metric
solution of GR equations -  remains a major issue of black hole
astrophysics. In this context, several effects are taken into account
for X-ray spectroscopy. Strong redshift of photons forming the
observed red wings, at $E<4$ keV,   is considered as evidence of rapid
rotation of a black hole (e.g., Brenneman \&  Reynolds 2006). However,
similar redshift can be obtained for $a=0$ if emission from $r_{\rm
d}<6$  is taken into account (Reynolds \& Begelman 1997).  Then, the
derived high value of $a$ relies on assumption of no neutral iron
emission  from within the radius of marginal stability.

Response of the line to increase of primary emission has been
suggested for future studies.  In particular, Reynolds et al.\ (1999)
indicate a bump occurring in the line profile and  progressing to
lower energies, with proceeding time, as a feature characteristic for
a rapidly rotating black hole. Again, a similar - redshifted and
Shapiro delayed - bump should appear  for a non-rotating black hole if
fluorescence inside $r_{\rm ms}$ was taken into account.

A straightforward analysis of the space-time metric could be performed
for systems observed close to edge-on, for which effects due to
lensing by a black hole would be directly seen in the line profile
(e.g., Zakharov \& Repin 2003).  However, as noted in Narayan \&
McClintock (2005), there seems to be  a selection effect preventing
such systems from being observed.

Then, we point out that a (largely unambiguous) analysis of imprints
of  the space-time metric would be possible in profile of the line
resulting from  irradiation by a strong flare just above the disc
surface.  If such a flare occurred at $r_{\rm s} \la 4$, properties of
the space-time related with  the black hole rotation would result in
$\ga 10$ per cent in magnitude effects in the line profile. The
effects related to the value of $a$ are rather subtle but in principle
possible to establish observationally.   A compact flare dominating
total emission would be required to make such an analysis feasible and
viability of such scenario is uncertain.  Such flares are indeed
occasionally observed (e.g., Ponti et al.\ 2004).  Interestingly,
however, the Fe K$\alpha$ line was actually very weak during the flare
analysed by Ponti et al.\ (2004), and a strong line appeared in the
spectrum with significant time delay after the flare.

Finally, we emphasise that the reduced variability of reflected
component -  basing on mechanism  advocated in this paper - is itself
a direct manifestation of the nature of the Kerr space-time. Note that
another effect resulting from  properties of the Kerr metric,  namely
anisotropic emission of hard X-rays generated close to a rotating
black hole, is qualitatively consistent with inclination-angle
dependence of intrinsic spectra of Seyfert galaxies (Nied\'zwiecki
2005).  In support for the tentative relation of these two effects to
properties of the Kerr metric, note also that comparison of the total
mass in black holes in the local  universe with the total luminosity
produced by active galactic nuclei indicates that most supermassive
black holes should rotate rapidly, e.g.\  Elvis, Risaliti \& Zamorani
(2002).

\subsection{Modelling the red wing of relativistic Fe lines}

As discussed above, detection of photons with $E < 4$ keV is
considered  as the evidence of rapid rotation of the black hole. Such
strongly redshifted photons were revealed in the Fe line profiles
observed in MCG--6-30-15 and several other objects (e.g., Miller et
al.\ 2004, Miniutti et al.\ 2004).   When fitted by models assuming a
power-law radial emissivity, the observed profiles require $q>4$
within $(6-10) R_{\rm g}$.  Considering physical scenarios for
generation of such profiles, we find that they
\newline
(i) can be produced   as a result of illumination by a hard X-ray
source located close to a black hole (preferably at $r_{\rm s} =2$--3)
and rather close to the disc surface ($h_{\rm s} \la 1$);
\newline
(ii) cannot be explained by models involving a source at a height of
several   $R_{\rm g}$, or higher, close to the axis;
\newline
if they are to come from a disc, with small inclination, surrounding a
rotating black hole. Regarding (ii) we find, in agreement with
previous studies, that such location of the source yields a  steep
$\epsilon_{\rm Fe}$, with $q>3$, in the innermost part of the disc.
However, this steep $\epsilon_{\rm Fe}$ occurs only within $r_{\rm d}
\le 2$ (and due to gravitational blueshift rather than light bending).
The majority of emission from this region of the disc is either
captured or bent toward the disc plane; moreover,  the observed
photons are strongly redshifted. As a result, this emission gives only
a minor contribution to the line profile and it is not relevant to the
shape and strength of the observed red wings.  This conclusion is not
affected by effects related with azimuthal motion of the source,
returning radiation or angular emission law of fluorescence; these
effects influence mostly the strength of the blue peak. On the other
hand, the red wing is primarily related with position of the X-ray
source.

\subsection{Returning radiation, non-local irradiation}

A further challenge for modelling certain line profiles results from
illumination of the surrounding disc (i.e. beyond a few $R_{\rm g}$)
by returning radiation. We find that this effect may be strong  for
the range of parameters (small $r_{\rm s}$ and $V$) previously not
explored.  A similarly strong enhancement of Compton reflection by
returning radiation, for very small distances ($r_{\rm s} \la 2$) of
primary source, was noted recently by Suebsewong et al.\ (2006).
 
Another effect, related to bending of photon trajectories to the disc
plane,  affects models relating the Fe radial profile to some physical
processes.  These models typically assume that the radial profile of
Fe emission is the same as some physically motivated profile of energy
dissipation.  E.g., Reynolds et al.\ (2004) make such assumption in
their analysis of the {\it XMM} observation of MCG--6-30-15, applying
model of the torqued-disc emission (where emission results mostly from
extraction of rotational energy of the black hole and thus is very
centrally concentrated). Moreover, they assume that there is no
dissipation, and thus no reprocessing, beyond rather  small (several
$R_{\rm g}$) outer radius.  We emphasise that even for X-rays
generated very low above the disc surface,  a significant illumination
of more distant regions of the disc  should occur.

Both the returning radiation and the non-locality of irradiation
result primarily in enhancement  of the blue peak of the line.

\subsection{Azimuthal motion}

Impact of strong gravity is usually studied under specific assumptions
on the X-ray source motion. Typically, angular velocity of the source
is related to that of the disc (e.g. Ruszkowski 2000; Miniutti \&
Fabian 2004), while Dabrowski \& Lasenby (2001) assume a static
primary source.  We note that change of $V$ significantly affects flux
and profile of the iron line through combination of SR and GR
effects. Moreover, the Kerr metric terms yield a non-trivial
dependence  between $V$ and $\Omega$ (equation (\ref{v})). Then, the
assumed parametrisation  of motion may be crucial for derived
properties, e.g.\ in some variability models.  Obviously, additional
effects may result from relativistic vertical or radial motion of the
source (e.g.\ Yu \& Lu 2001), which were neglected in this paper.

\subsection{Applicability of our results}

We focused above on MCG--6-30-15, for which several observed
properties may be explained by our model. Similar effects, including
reduced variability of reflection, pronounced red wings or strongly
reflection dominated spectra, have been revealed in some Narrow Line
Seyfert 1 galaxies (e.g. Fabian et al.\ 2004, Ponti et al.\ 2006), as
well as in stellar mass black-hole systems (e.g.\ Miniutti et al.\
2004) and in galactic nuclei with much higher  black hole masses
(Fabian et al.\ 2005).  Note that our scenario implies that the X-ray
luminosity   is underestimated by an order of magnitude.  in these
objects.

On the other hand, most black-hole systems do not show signatures  of
such extreme effects. Then, similar reduction of the X-ray flux does
not necessarily characterise  any low inclination system.  Note,
however that such property would be inevitable for rapidly rotating
black holes, where major fraction  of accretion power is dissipated at
very small  $r_{\rm s}$, if this power is converted  into hard X-rays
in situ.

As in virtually all previous studies of that subject, we analysed
impact of GR  effects for the simplest case of an isotropic point
source of hard X-ray emission, which approach follows from our lack of
understanding of the X-ray source nature.  The derived properties
result from transfer of radiation from source to disc and observer and
from disc to observer.  Additional (but smaller in magnitude) effects
may occur in more realistic models.  E.g., Comptonization close to a
Kerr black hole gives rise to anisotropic emission, cf.~Nied\'zwiecki
(2005), therefore radiation with different spectral index  may
irradiate various parts of the disc,  while the transfer effects
affect only normalisation and cut-off energy of the primary emission.

The simplified description of X-ray source considered in this paper
approximates  most closely scenario with magnetic flares above the
disc surface. Then, our results may be directly applicable to a model
with flares occurring randomly at various radial distances.
Qualitatively, we may assess similar variability effects  for
continuous spacial distributions of the hard X-ray source, e.g.\ an
extended corona covering the disc surface or a small hot torus
replacing the disc within a few innermost $R_{\rm g}$. Varying size of
such an extended, hot plasma, in the Kerr metric, should give rise
reduced variability of reflected component. In particular, a decline
of direct emission from a  shrinking corona or torus would be
observed, even if its intrinsic luminosity remained unchanged, while
increasing fraction of its emission would be bent to the disc plane
giving rise to strong reflection component.

The major shortcoming in our study results from the neglect of
ionization effects. The strongest ionization should occur below the
source, especially in models with low height above the disc surface.
In general, strong ionization of this region should suppress the
redshifted Doppler horns formed in the red-wing.  On the one hand,
this would make studies of effects related to value of $a$ less
feasible. On the other hand, such depletion of the variable
contribution to the red wing would reduce variations of the line
profile at various flux states.  However, details of ionization
structure, and of the related reduction of the variable contribution
to the lien, depend on additional assumptions, in particular, on
azimuthal distribution of flares. Obviously, a strong single flare
would give rise to stronger ionization  than  many weak flares
uniformly distributed at a given $r_{\rm s}$.

\subsection{Summary}

If attributed to strong-gravity effects, both the time-averaged line
profile  and the variability pattern observed  in MCG--6-30-15 by {\it
XMM-Newton\/}  independently indicate that a primary  hard X-ray
source must be located very close  ($\la 4R_{\rm g}$) to a black hole,
i.e.\ in the region where Kerr metric effects  become crucial.  Rapid
rotation of the black hole is necessary (and vastly sufficient) to
account for  reduction of variability of reflected emission for such
spacial location of the  source.  Bending to the equatorial plane,
underlying this reduction effect,  appears to be the most pronounced
effect  of the Kerr metric to be studied in the X-ray spectra of
black-hole systems.

\section*{Acknowledgments}
We thank the anonymous referee for detailed comments and suggestions
to improve the paper.  This work was partly supported by grant
no. 2P03D01225 and  N203 011 32/1518 from the Polish Ministry of
Science and Higher Education.

\label{lastpage}
\end{document}